# Mie Sensing with Neural Networks: Recognition of Nano-Object Parameters, the Invisibility Point, and Restricted Models

*Artur Movsesyan, Lucas V. Besteiro, Zhiming Wang,\* and Alexander O. Govorov\**

A. Movsesyan, Z. Wang,
Institute of Fundamental and Frontier Sciences, University of Electronic Science and Technology of China, Chengdu 610054, China
E-mail: zhmwang@uestc.edu.cn
A. Movsesyan, A.O. Govorov,
Department of Physics and Astronomy, Ohio University, Athens OH 45701, USA
E-mail: govorov@ohio.edu
L.V. Besteiro,
CINBIO, Universidade de Vigo, 36310 Vigo, Spain

In this work, we use artificial neural networks (ANNs) to recognize the material composition, sizes of nanoparticles and their concentrations in different media with high accuracy, solely from the absorbance spectrum of a macroscopic sample. We construct ANNs operating in the following two schemes. The first scheme is designed to recognize the dimensions and refractive indices of dielectric scatterers in mixed ensembles. The second ANNs model simultaneously recognizes the dimensions of gold nanospheres in a mixture and the refractive index of a matrix. A challenge in the first scheme arises at and near the invisibility point, i.e., when the refractive index of nanoparticles is close to that of the medium. Of course, particle recognition in this regime faces fundamental physical limitations. However, such recognition near the invisibility point is possible, and our study reveals its unique properties. Interestingly, the recognition process for the refractive index in the vicinity of the invisibility point shows very small errors. In contrast, the errors for the recognition of the radius grow strongly near this point. Another regime with limited recognition occurs when the extinction spectra are not unique and can correspond to different realizations of nanoparticle mixtures. Regarding multi-particle or polydisperse solutions, the ML-based models should in such cases be rationally restricted to maintain the feasibility of the recognition process. Overall, the recognition schemes proposed and investigated by us can find their applications in the field of sensing.

## 1. Introduction

Nano-objects can resonantly scatter light in the UV-VIS-IR region due to the excitation of the Mie resonances[1–3] and the localized surface plasmons,[4–6] respectively in dielectric materials and metals. These optical resonances depend on the nano-object material, shape, and size. [7,8] There is a wide variety of analytical and numerical tools for the computation and analysis of the electromagnetic response of nano-objects, such as the Mie theory,[8] T-matrix method,[9] Discrete Dipole Approximation (DDA), Boundary Element Method (BEM), Finite Element Method (FEM), and Finite-Difference-Time-Domain approach (FDTD).[10] The aforementioned methods provide solutions for the direct problem (to determine the optical spectra from a given material, shape, and size) for virtually arbitrary cases (**Figure 1a**), but the inverse problem (to define the physical parameters of the nano-object from the spectrum, see Figure 1b, 1c) remains a challenging task because it requires sampling a significant section of the design space each time. Overcoming this difficulty and developing useful general methods to solve the inverse problem in nanophotonics is of great interest because it would allow the rapid design of systems with unique optoelectronic properties. Thus, this area has attracted much interest for at least the last two decades, and several new approaches based on the transformation optics,[11] and optimization techniques,[12–16] were proposed.

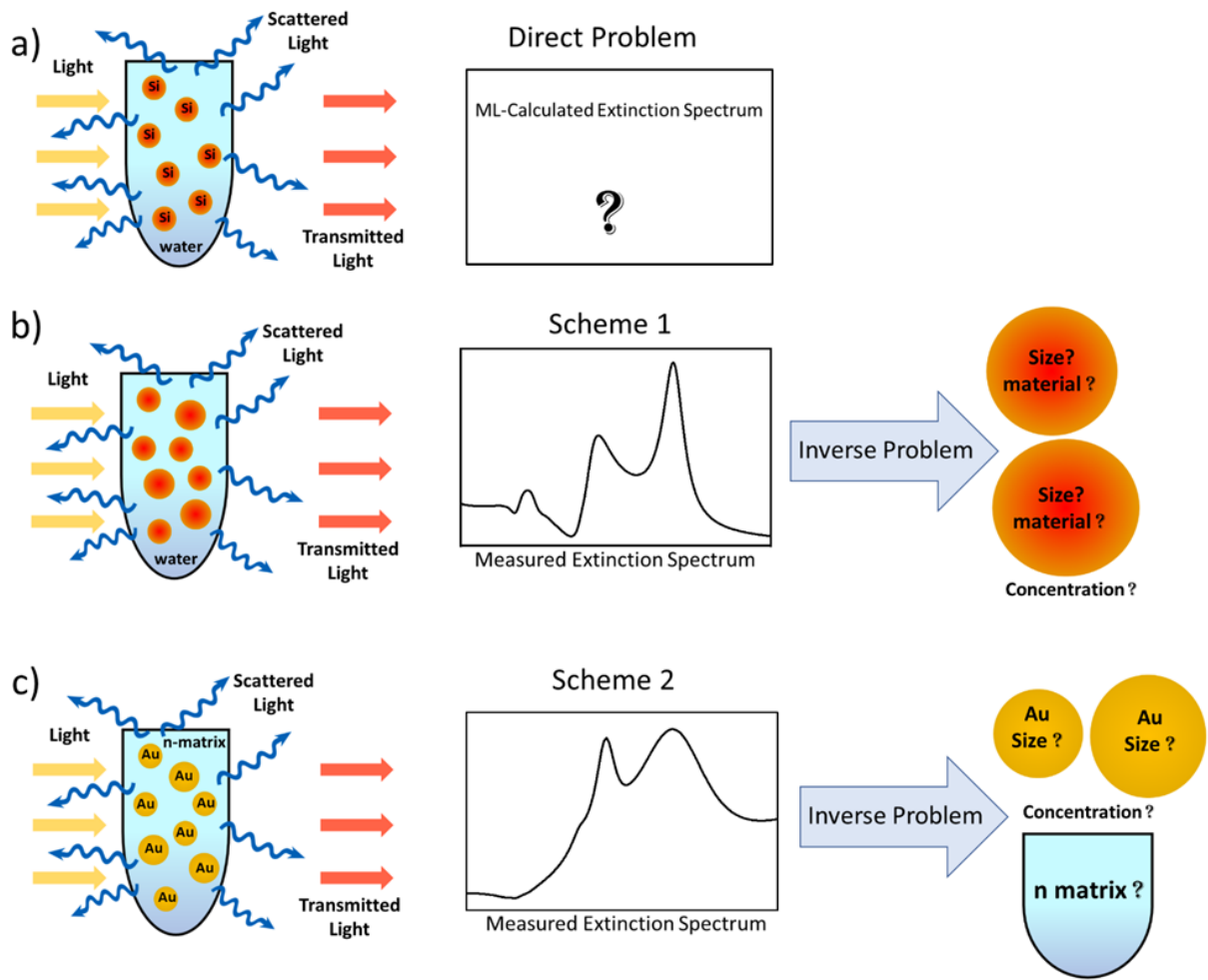

**Figure 1.** Schematic description of the direct (a), and the inverse problem (b,c). The direct problem consists of determining the optical extinction spectrum of a known system composed of solvent and nanostructure ensemble. The inverse problem considers the determination of (b) the properties of an unknown nanostructure from a given scattering spectrum, e.g., sizes, refractive indexes and their concentrations, in our Scheme 1, and (c) the sizes of gold nanostructures, their concentrations, and the refractive index of an unknown dielectric matrix where the nanostructures are embedded, for Scheme 2.

During this time, advances in computer hardware have enabled the implementation of ideas developed in the field of artificial intelligence during the last century, leading to impressive demonstrations of the potential of artificial neural networks (**ANNs**) in machine learning. These techniques have been adopted into a variety of fields in the natural sciences, including for solving inverse problems in nanophotonics[15,17,18] and also in nanospectroscopy,[19] material sciences,[20,21] and microscopy.[22,23] Typically, supervised learning employing ANNs is used in two main procedures aimed at solving the inverse problem: classification[19,24] and regression.[25–28] In the majority of studies considering computational inverse designs in nanophotonics and its subdomains, ANN architectures are used to formulate a regression problem for the prediction of physical parameters capable of producing the desired response. Those parameters can be, e.g., nanostructure dimension,[25,26,29] morphology[30] and shape,[23] or chiral nano-object design.[27–29] Nonetheless, there are notable works showing interest in performing classification tasks in nanophotonics, e.g., distinguishing single-photon quantum emitters,[31] strengthening readout in high-density data storage,[19] or the label-free identification of pathogenic bacteria[24] and cells.[32] The inverse problem can be addressed also through a mixed approach, e.g., when the physical dimensions of nano-objects are predicted by regression, while their material is labeled and classified.[29] Additional frameworks, such as semi-supervised or unsupervised training of neural networks, e.g., using generative adversarial networks, are also used for solving inverse problems in the engineering of metamaterials[33,34] and material discovery.[21] Along with treating purely optical responses, the ML approaches and recognition formalisms can be applied in the research fields related to the hot-electron effects[35] and photothermal phenomena.[36]

Nowadays, neural networks have found a large number of applications, for example, natural language processing,[37] image recognition,[38,39] super resolution microscopy,[40,41] design and optimization of nanophotonic devices.[26,42,43] However, the recognition of the nano-objects in solution remains a challenging problem with a relevant history in nanophotonics (see section 1.5 in Ref. [44]), due to the very large parameter space that would require being sampled to solve it in general. In this letter, we address some of the central subsets of the general recognition problem, discussing the fundamental physical challenges that they pose, and considering the implementation of its solution using ANNs, an approach that is especially well-suited for running in low-power devices for, e.g., off-grid measuring stations. In our study, we formulate two types of recognition schemes for optically active nano-objects: (1) Simultaneous determination of the sizes, dielectric constants, and concentrations of a mixture of dielectric scatterers made of an unknown material embedded in a known medium; (2) Simultaneous recognition of the refractive index (**RI**) of the local environment, the radii of plasmonic nanoparticles, and their concentrations. In this study, we will take advantage of deep neural networks to predict and recognize certain as similar as well as dissimilar nano-objects with high precision, above 99%, in wide intervals of the nanosystem's parameters. At the same time, we show that the recognition process based on Machine Learning (ML) has fundamental limitations. In particular, challenges in the recognition process appear at and in the vicinity of the optical invisibility point when the RI of a nano-object is close to that of a matrix. We show that the choice of the ML model, the involved material systems, and the details of the training (for example, the intervals of parameters) become crucial for the recognition in the above-mentioned regime. The recognition near the invisibility point is possible and, interestingly, the errors for the refractive index in that regime are low. Simultaneously, the determination process for the nanoparticle size generates large errors. We found the above behaviors for both one- and two-component solutions. Another limitation and challenge, which we identify in this study, concern the cases with physical degeneracies in the scattering spectra when different systems exhibit very similar optical spectra.

Overall, the recognition process becomes more challenging for a small nanoparticle size, whereas the recognition accuracy greatly improves for larger spheres with well-developed Mie resonances. Finally, we note that our approaches can be used for various material systems, including mixtures of plasmonic and dielectric nanoparticles. In mixtures with a few or many nanoparticles' sizes, the number of parameters to recognize grows rapidly with the number of fractions. Of course, the problem of polydisperse or many-component solution in the general case represents a real challenge. Then to make the recognition scheme practical, a many-component model should be rationally restricted; otherwise, the ML procedure becomes computationally infeasible. In the following, we demonstrate the recognition schemes for various one- and two-component mixtures with two-five parameter sets. The number of components in a mixture can potentially be further increased.

The manuscript is constructed by first discussing a relatively simple case for a given recognition scheme (two parameters) and then analyzing the limitations and challenges in each section. After that, we investigate the cases with more than two parameters.

## 2. Recognition of Nano-objects.

### 2.1 Formalisms

To achieve an effective recognition of sizes and refractive indexes of nanospheres (**NSs**) from their optical response, we employed a well-established ANN model (multilayer perception operating as a regressor), and implemented it using the Scikit-learn machine learning library in Python (**Figure 2a**).[45] The datasets of spectra used for the training, validation, and testing of our ANNs were prepared with the help of Mie Theory (Supporting Information). The training, validation, and testing datasets of the Mie spectra (or the corresponding parameter sets) do not overlap, of course. In our study, we took 75% of our total Mie spectra for training (90% of 75%) and validation (10% of 75 %). Correspondingly, the set for the testing included the remaining 25% of the total spectra (See more details for all ANNs in Table S1 in the SI). The logic of the paper is the following. We start with the simplest two-parameter problem (**Figures 3,4**), which also incorporate the invisibility point $(n = n_m)$, where $n$ and $n_m$ are the refractive indices of NSs and matrix, respectively. Then, we increase the number of the recognition parameters and also consider the plasmonic case. Considering the case of a mixture, we follow the typical experimental scheme in the optical spectroscopy of colloids and aerosols that is based on the Beer-Lambert law.

$$T = 10^{-OD}, \qquad (1)$$

where $T$ is the transmission through a NS solution and $OD$ is the optical density.

$$OD(\lambda) = \frac{L_{opt}}{\ln(10)} \sum_i \sigma_i(\lambda) \cdot \rho_i, \qquad (2)$$

where $L_{opt}$ is the optical path that the light traverses through the NS solution. The sum index runs through the different types of NSs in an ensemble, and $\rho_i$ and $\sigma_i$ are the number concentration and the extinction cross-section of the $i$-th species, respectively. In our paper, we fix the optical path to be $L_{opt} = 1$ cm. Since the $OD(\lambda)$ parameter can be measured experimentally, it is logical to use in the ML recognition procedure the following spectral function:

$$f(\lambda) = \sum_i \sigma_i \rho_i.$$

**Figures 5,7 and 8** shows the results where the $f(\lambda)$ function is used to recognize/sense the unknown solution parameters such as $n_i, R_i, n_m$, and $\rho$; here $R_i$ is the radius of the *i*-NS in the mixture. Table 1 below summarizes the set of the ML tasks accomplished here. The numbers of parameters in our study are in the range of 2-5; the two- and three-parameter cases are not challenging computations but serve as an important demonstration of the principle and as the starting point for the recognition approach. We note that, in the current literature, one can see ML data for optical tasks with 3-8 parameters.[25,27–29,46–48]

**Table 1.** The left column shows the number of recognized parameters. The central column shows the recognized physical parameters, where $n$ is the refractive index of NSs, $R$ is the radius of NSs, $\rho$ is the concentration (number of particles per cm³) of NSs in solutions, $n_\mathrm{m}$ is the refractive index of the medium. $\alpha$ is relative concentration of NSs ($R_2$).

| Dielectric Nanoparticles | | |
|---|---|---|
| 2 parameters | $(n, R)$ | Figure 3, 4, S2 |
| 3 parameters | $(n, R, \rho)$ | Figure 5, S3, S4 |
| 3 parameters | $(n, R_1, R_2)$ | Figure S5 |
| 5 parameters | $(n_1, R_1, n_2, R_2, \alpha)$ | Figure 6, S6 |
| **Plasmonic Nanoparticles** | | |
| 2 parameters | $(n_\mathrm{m}, \rho)$ | Figure 7 |
| 2 parameters | $(n_\mathrm{m}, R)$ | Figure S7 |
| 3 parameters | $(n_\mathrm{m}, R, \rho)$ | Figure 8, S8 |
| 4 parameters | $(n_\mathrm{m}, R_1, R_2, \alpha)$ | Figure S9 |

## 2.2 Recognition of Mie-nanospheres and their mixtures

*2.2.1 The basic two- and three-parameter schemes for Mie nanospheres with and without the invisibility point*

Firstly, we present the results obtained with ANN₁ and ANN₂ for the recognition of two parameters, e.g., the radii and refractive indices of NSs embedded respectively in air and water. The first network, ANN₁, is adapted for the recognition of NSs with radii in the interval between 30 nm to 100 nm, and with refractive indexes from 1.35 to 5. Figure 2b shows the optical spectra of the two nanospheres and the predicted physical parameters. To evaluate a single object recognition, one may use the relative error for a predicted value,

$$Error_\mathrm{Rel} = \frac{|y_\mathrm{true} - y_\mathrm{pred}|}{y_\mathrm{true}}$$

In the case of NS of $R=30\,\mathrm{nm}$ and $n=1.35$, values at the extremes of the training dataset, the error is 0.004 for the radius and 0.059 for the RI. For the NS of $R=100\,\mathrm{nm}$ and $n=5$, the relative error is 0.005 for radius and is 0.01 for the RI. The Mie scattering of with disparate values of RI and radius can produce distinct spectral forms with non-resonant, mono-resonant, or multi-resonant optical excitations. These features one may find in Figure 2d,e respectively for NSs in air and in water.

It is worth summarizing several major difficulties that we may face while identifying NSs across the range of parameters studied. These are fundamental physical limitations that introduce spectral degeneracy between NSs and will thus limit the precision of our numerical scheme. Mainly, these arise when considering particles with small sizes, and refractive indexes that are small or close to that of the environment. These conditions lead to low scattering cross sections or resonances uniquely in the Vacuum-UV to Mid-UV (50-300 nm), which is beyond the chosen spectral range. Consequently, the ANN has to distinguish systems due to small differences in UV-VIS spectra, or only showing an exponential decay mode (black curve Figure 2b). While the large and high index NSs show resonant multipolar excitations. This feature is reflected on the spectrum of the NS with $R=100\,\text{nm}$ and $n=5$ (Figure 2b). The peaks present the excited magnetic and electric optical modes in the NS.

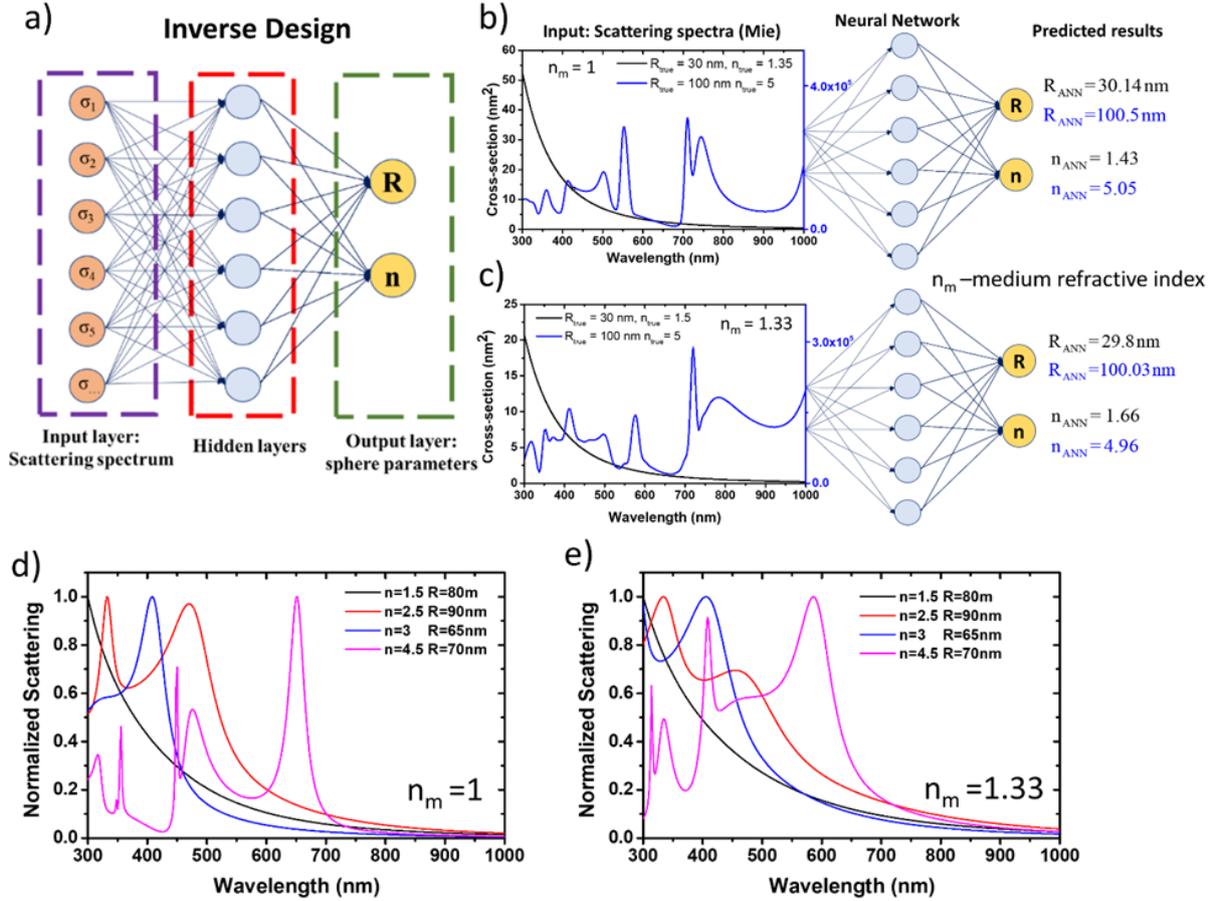

**Figure 2.** (a) Diagram of the ANN considered for the inverse problem. The input parameters of the ANN are the optical scattering spectra of nanospheres, and the output parameters are the radius and the refractive index (RI) of the nanospheres. The input layer has 71 neurons that correspond to the number of datapoints of the optical scattering spectrum. The output layer has 2 neurons which correspond to two predicted physical parameters, i.e. the radius ($R$) and the RI ($n$) of the NSs. The ANN has 4 hidden layers with the following configuration of neurons: 200-500-200-30 (see details of ANN in Supporting Information). One may find the results for other configurations of ANNs in Table S2 (Supporting Information). (b, c) Optical scattering spectra of nanospheres, the predicted sizes and refractive indexes embedded in (b) air ($n_m=1$) and (c) water ($n_m=1.33$). Examples of normalized Mie scattering spectra of different refractive indexes ($n$) and radii ($R$) nanospheres (d) in air and (e) in water.

The accuracy of the prediction of refractive index is much less for NSs of the low refractive index of small radius compared with the NS with the high RI. However, the accuracy of the performance of ANN$_1$ is evaluated by $r^2_{Score}$. We confirm the high goodness of the predicted results with expected ones since ANN$_1$ shows a high prediction accuracy ($r^2_{Score}=0.9993$).

Figure 2c shows the optical scattering spectra of NSs submerged in water and, correspondingly, the predicted sizes and refractive indexes. ANN$_2$ is adapted for the prediction of NSs radii in the interval between 30 nm and 100 nm, and refractive indexes from 1.5 to 5. Both ANN$_1$ and ANN$_2$ operate out of the invisibility-point regime $(n \neq n_\mathrm{m})$. One may note that with ANN$_2$, the predicted value for the RI of small NS has a larger deviation than the air case, in ANN$_1$ reaching relative errors of up to 0.106. The ratio between refractive indexes NS/environment plays an important role in the prediction accuracy. The prediction accuracy drops when the ratio approaches 1, thus approaching the invisibility point. However, the overall prediction ability is $r^2_\mathrm{Score} = 0.9985$ in this case. Consequently, with both ANN$_1$ and ANN$_2$ we achieve the recognition of two physical measures with a different nature and scales of magnitude, i.e., the NS dimension and RI, with a high coefficient of determination $(r^2_\mathrm{Score} > 0.99)$.

To analyze and discuss the recognition skills of ANNs for a wide range of different NSs, we present in Figure 3 the relative errors of the predicted results of refractive indexes and radii. Figure 3a and 3b shows the relative error of the predicted physical parameters of NSs embedded in air $(n_\mathrm{m} = 1)$. The largest relative errors for RI prediction are found for the smallest NSs $(R = 30\,\mathrm{nm})$. Overall on the maps of relative errors for the predicted radii, the highest uncertainty may be found on the bottom-left corner, for NSs of small sizes and low refractive indices.

Figure[s] 3c and 3d presents the relative error maps for NSs embedded in water. Similar to the previous case (air medium), the highest errors for the refractive-index prediction are found for the NSs with small sizes and low refractive indices. This tendency can be attributed to the above-mentioned major difficulties, i.e., very small scattering cross-sections and resonances in the Vacuum-UV to Mid-UV. To clarify this behavior, we computed the maps of the mean of scattering spectra that highlight the regions of low scattering (see **Figure S1** in the SI). One may note that the maximum values on the maps of relative error correspond to the regions with low scattering amplitudes in Figure S1.

To emphasize the challenges and limitations that ANNs may face in the recognition of NSs within the fixed interval of these two physical parameters, we show the relative error for three selected radii in Figure 3e, f. We also introduce a threshold of acceptability at a relative error equal to 0.05. One may observe that, unlike with the small NSs and those with low RI, the recognition errors are mostly well below that threshold value in our recognition scheme.

Next, we look at more complex cases, where the prediction ability of our ANNs (ANN$_3$ and ANN$_4$) are tested in the vicinity of the invisibility point $(n \approx n_\text{m})$. We extended the interval of considered RI for the NSs for both air and water media ($1 < n < 5$, $30\,\text{nm} < R < 100\,\text{nm}$). The prediction scores ($r^2_\text{Score}$) of ANN$_3$ and ANN$_4$ (Figure 4) for this interval of the refractive indexes are $r^2_\text{Score} = 0.9977$ and $r^2_\text{Score} = 0.9854$, respectively, for the air and water media. We see that the prediction accuracy drops compared to previous cases, where we did not consider the invisibility point regime.

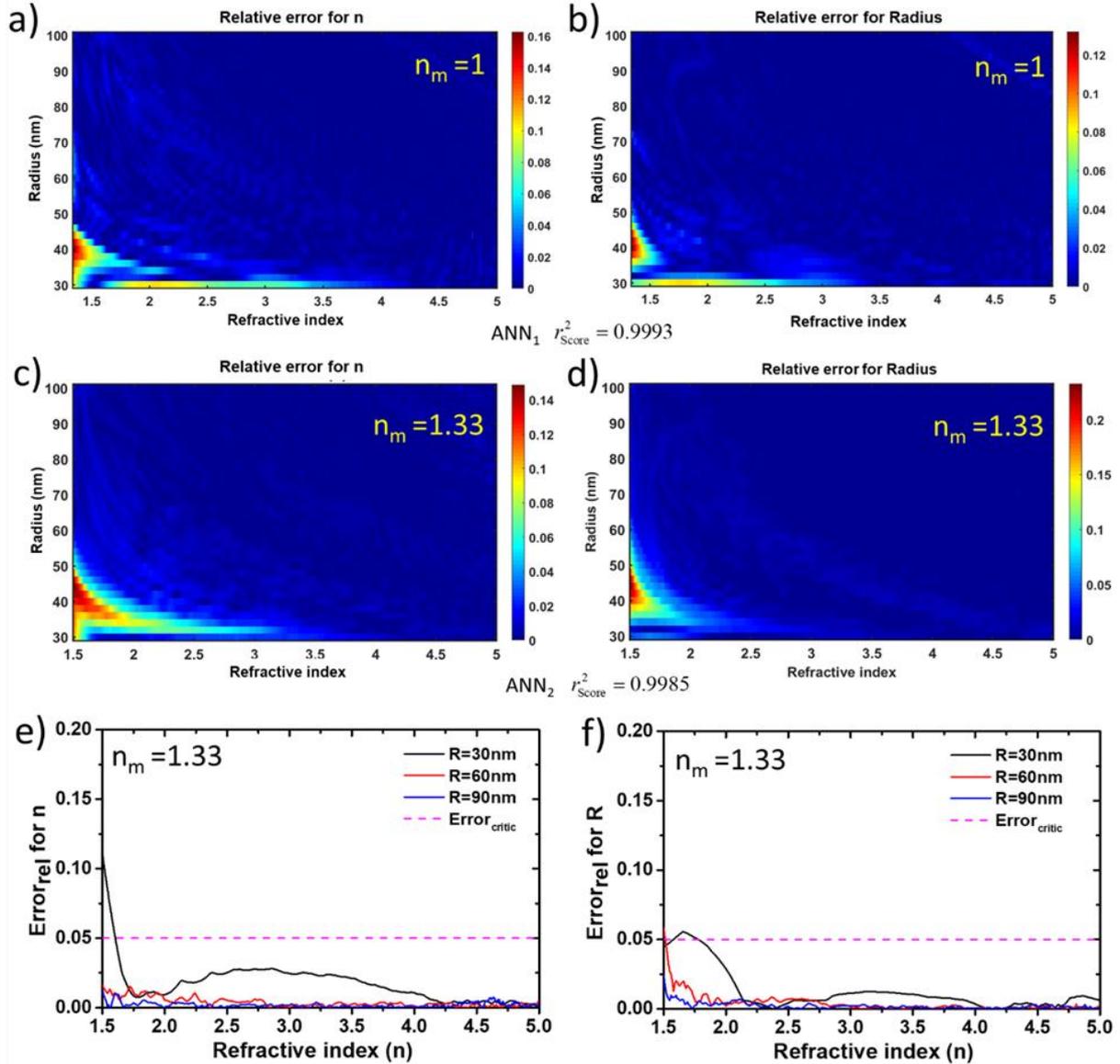

**Figure 3.** Two-parameter recognition scheme. The figure depicts the relative-error maps for the refractive-index and radius prediction obtained from the two trained ANNs for two matrix media (air and water). (a,b) Nanospheres in air: $n_m=1$ and $1.35<n<5$. (c,d) Nanospheres in water: $n_m=1.33$ and $1.5<n<5$. (e,f) Relative-error curves of the recognition procedure for the RI and the radius for three selected cases: $R=30$ nm, 60 nm, 90 nm. Local environment refractive index is $n_m=1.33$. The dashed line highlights the barrier of critical relative error ($Error_{Critic}$). We see that, except for some particular intervals of the parameters in panels (e,f) the relative error is below the chosen critical value. As expected, the maximum errors occur for the smallest size of 30 nm.

Figure 4a depicts the map of relative error for the predicted refractive indexes of nanospheres ($30\,\text{nm}<R<100\,\text{nm}$ and $1<n<5$) in air. The highest inaccuracy in the prediction of the RI occurs for the small and low-refractive index NSs. We drew the relative error map of the predicted radii in Figure 4b. In the figure, it is particularly noticeable the inaccuracy of the prediction of the radii near the special (i.e., invisibility) point $(n \approx n_m = 1)$.

Next, we present the relative-error maps for the refractive indices of NSs $(30\,\text{nm} < R < 100\,\text{nm}$ and $1 < n < 5)$ placed in water $(n_m = 1.33)$ in Figure 4c. Interestingly, the highest inaccuracy of the prediction of refractive indices is not at the invisibility point; but right outside of it. Unlike the refractive-index relative errors, the highest relative errors of the predicted radii are near the invisibility point (Figure 4d) since a NS of any size has zero scattering cross-section in this regime. Simultaneously, considering the recognition of the RI in Figure 4c, the RI may be correctly set to $n \approx n_m$ by an ANN even for zero input value of the scattering cross-section. The big error of the predicted radii at the point $n = 1.33$ in Figure 4f conceals the errors on the left $(n < 1.33)$ and right $(1.33 < n < 1.7)$ sides of the special invisibility point, which are clearly visible on the error map of refractive indices (Figure 4c).

To highlight the challenge posed by these regimes, we also present the curves of relative error for three selected radii ($(R=30\,\text{nm}, 60\,\text{nm}, 90\,\text{nm})$) in Figure 4e,f. It is clear that the value $n = 1.33$ is a special, unviable point for our recognition scheme when a nano-object is invisible. However, the NSs can be recognized with small $n$-errors in the vicinity of this special point $n = 1.33$ $(1.3 < n < 1.35)$, as one can see in the inset of Figure 4e. Simultaneously, near this special point, the ANN shows very high values for the $R$-errors (Figure 4f and its inset).

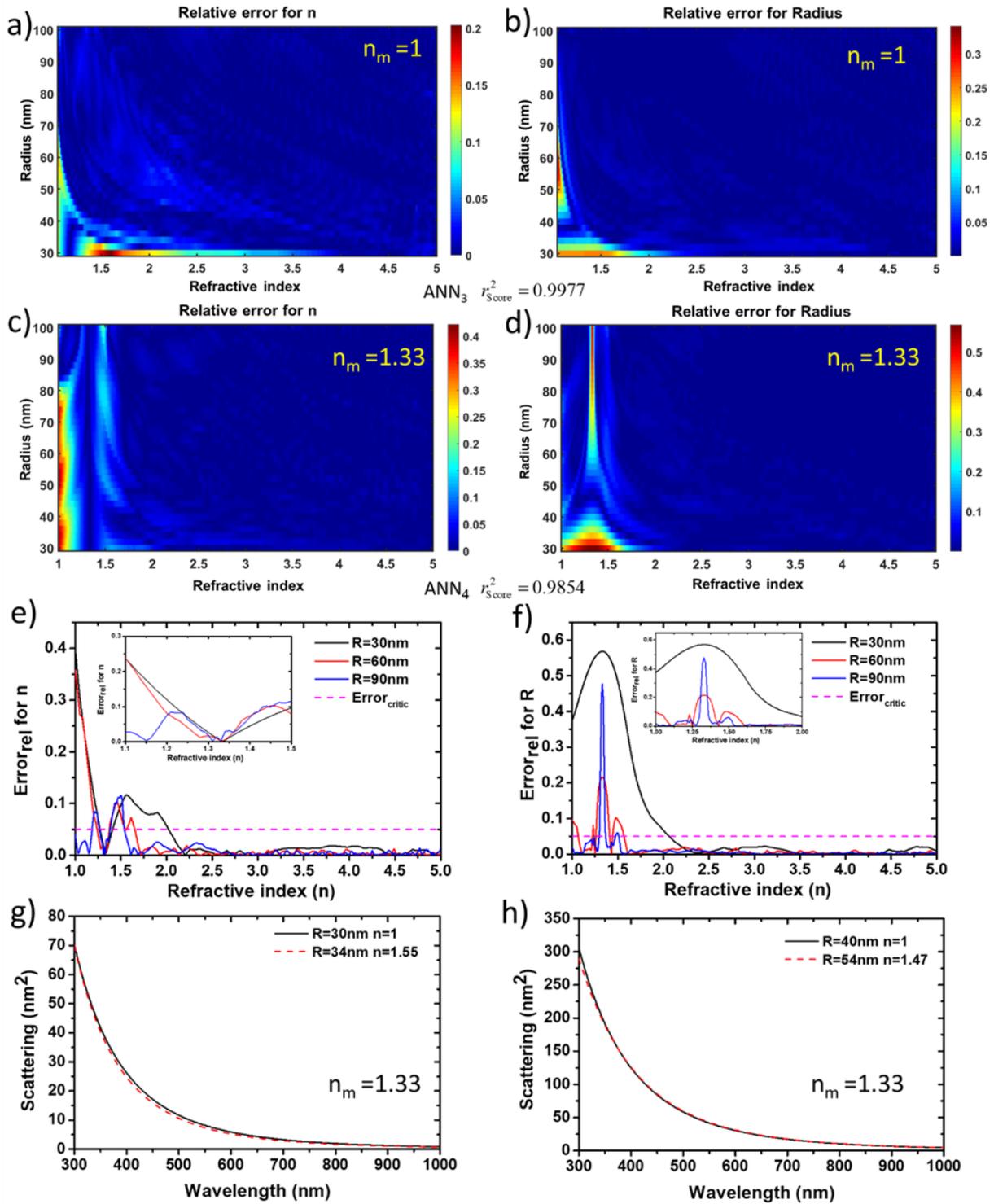

**Figure 4.** Two-parameter recognition scheme involving the invisibility point. We show here the relative-error maps for the refractive-index and radius prediction obtained from two trained ANNs for two matrix media (air and water). (a,b) Air matrix: $n_m=1$ and $1<n<5$. (c,d) Water matrix: $n_m=1.33$ and $1<n<5$. (e, f) Relative error curves of recognition of RI and radius for $R=30\,\text{nm}, 60\,\text{nm}, 90\,\text{nm}$. The local environment RI is $n_m=1.33$. The insets zoom into the region around $n=1.33$. (g, h) Examples of the nearly identical scattering spectra arising for two NSs with different

sizes and refractive indices. In a pair of NSs, the RI of one NS is below $n_\mathrm{m}$ $(n<1.33)$ and another one is above $n_\mathrm{m}$ $(n>1.33)$.

To understand this behavior of relatively big errors of the recognition in the intervals $n<1.33$ and $1.33<n<1.7$, we need to look into the data presented in Figure 4g,h. The NSs having a RI smaller $(n<1.33)$ and larger $(1.33<n<1.7)$ than the matrix parameter $(n_\mathrm{m})$ may show similar spectra. Here we should note that the Mie coefficients depend on the ratio of refractive indices. This results in the spectral degeneracy between different NSs when considering materials with refractive indices above and below $n_\mathrm{m}$.

Indeed, the ANN4 has inputs (spectra) similar to each other, while the outputs (radii and refractive indices) are different. This non-uniqueness of the inputs may cause the failure of the convergence of the ANN.[43] Although the errors appear next to the invisibility point due to the non-uniqueness of the inputs, overall we obtained a high accuracy for ANN4 $(r^2_\mathrm{Score}=0.9854)$. One may find the separate evaluation of the recognition in two regimes $(1<n<1.7$ and $1.7<n<5)$ in **Figure S2** (Supporting Information). It may seem that the recognition of two parameters is a simple task. However, the challenges arising at the invisibility point and its nearby make the recognition process nontrivial.

Then, we exploited our ANN model to recognize dielectric NSs of different concentrations $(\rho)$. We chose two intervals of refractive indices for NSs, where the first interval $(2.5<n<4)$ did not include the invisibility point in contrast to the second one $(1<n<2.5)$. The intervals of NSs densities are adjusted to obtain $0.01<OD<10$ (Equation 2) for the NS of $R=60\,\mathrm{nm}$ for refractive indexes $n=3$ and $n=2$, respectively, for the RI intervals $2.5<n<4$ and $1<n<2.5$.

For the first interval $(2.5 < n < 4)$, one may note that the addition of a new parameter $(\rho)$ to recognize did not influence much the recognition accuracy (Figure 5a,b) when we did not use challenging intervals. Figure 5c,d shows the relative error maps of the recognition for the refractive indices and radii for a particular span of the RI, $1 < n < 2.5$. The maps highlight the large errors near the invisibility point, analogically to the two-parameter recognition scheme by ANN₄ (Figure 4c,d).

The curves of relative error (Figure 5e,f) for three selected radii ($R$=50 nm, 70 nm, 90 nm) show that the refractive indexes $(n)$ can be identified with minor errors at the vicinity of the invisible point for the 3-parameter recognition scheme. On the contrary, the recognition of the radii of NSs is introduced with apparent errors. A similar behavior was observed for the two-parameter recognition scheme (See Figure 4e, f).

One may find 2D maps of relative errors for other concentrations of $\rho$ for the RI intervals of $2.5 < n < 4$ and $1 < n < 2.5$, respectively, in **Figure S3** and **Figure S4** (SI). These maps demonstrate that with the reduction of concentration, the errors become more evident. The decreasing of the concentration leads to the lowering of the amplitude of spectra. Hence, this decrease influences the accuracy of the recognition.

In addition, we employed our ANN model to recognize a mixture of two different NSs $(R_1, R_2, n)$ of similar concentrations (relative concertation, $\alpha = 0.5$) in water (See **Figure S5**, SI).

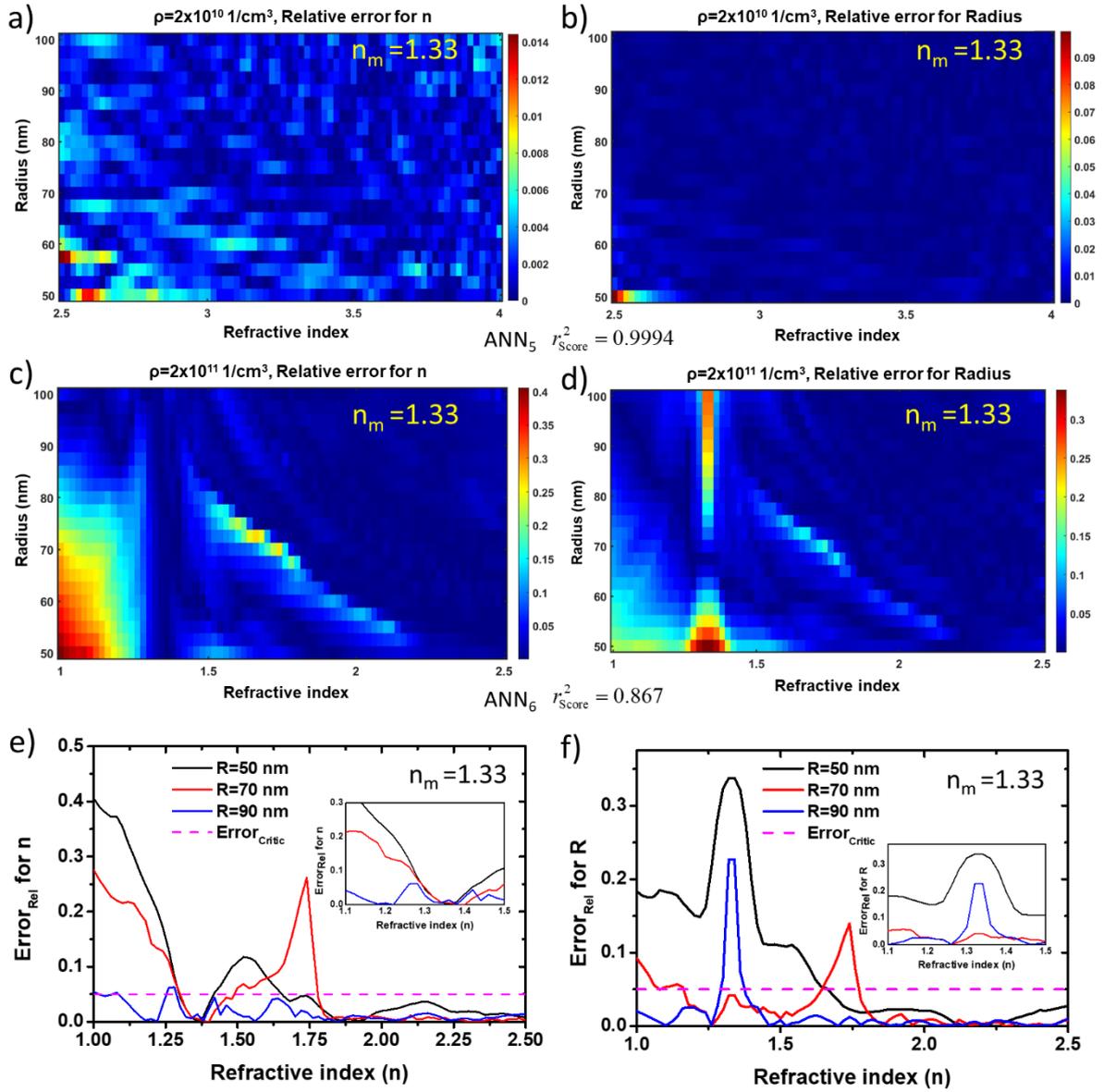

**Figure 5.** Three-parameter recognition scheme for NSs ($R, n, \rho$). The figure shows the relative-error maps for the refractive-index ($n$) and radius prediction ($R$) obtained from two trained ANNs for two intervals of refractive indexes: (a,b) $2.5 < n < 4$ and (c,d) $1 < n < 2.5$; the radii interval is $50\,\mathrm{nm} < R < 100\,\mathrm{nm}$. The density interval for (a,b) is $4*10^8\,1/\mathrm{cm}^3 < \rho < 4*10^{11}\,1/\mathrm{cm}^3$ and for (c,b) $2*10^9\,1/\mathrm{cm}^3 < \rho < 4*10^{12}\,1/\mathrm{cm}^3$. For the matrix, $n_\mathrm{m} = 1.33$. To make these 2D maps, we fixed one of the parameters, e.g., the density ($\rho$), since we deal with three-dimensional data. (e, f) Relative error curves of recognition of RI and radius for $R$=50 nm, 70 nm, 90 nm. The local environment RI is $n_\mathrm{m}$=1.33. The insets zoom into the region around $n$=1.33.

*2.2.2 The five-parameter scheme for a mixture.*

In the next step, we focused on a more global problem; namely, we aimed to employ the recognition scheme of two NSs by using just a single optical spectrum. These NSs may have different sizes, compositions, and concentrations. To accomplish this task, it is required a recognition of five parameters such as two radii, two refractive indices, and the relative concentration ($\alpha$) of one of the NSs. For this recognition scheme, the optical spectrum is defined using the following formalism:

$$\sigma_{\text{Total}}(\lambda) = (1-\alpha) * \sigma_{R_1,n_1}(\lambda) + \alpha * \sigma_{R_2,n_2}(\lambda),$$

where $\sigma_{R_1,n_1}, \sigma_{R_2,n_2}$ are the scattering cross-sections of two NSs, and $\alpha$ is the concertation coefficient of the second NS. We restricted the range of the parameters for the training of ANN$_7$ and ANN$_8$ to control the explosive/geometric increase of input spectra with the addition of a single parameter. The 2D maps of relative errors of the recognition of the RI ($n_2$) and radius ($R_2$) show small errors for the non-challenging regime (**Figure 6a,b**). One may find 2D cuts of relative errors for the other concentration than $\alpha = 0.6$ in SI (**Figure S6**). When the range of parameters does not include the invisibility point, the score of ANN$_7$ remains high, similar to the previous cases, despite the larger number of recognized parameters. In Figure 6e, we show a few selected examples of the recognition of two non-similar NSs and their relative concentrations. The resulting optical spectra based on the recognized parameters (red dashed line curves, Figure 6e) show good similarity with the ground truth spectra (black curves, Figure 6e). This observation let us conclude that our ANNs can run with high efficiency for both two- and five-parameter recognition schemes when the range does not include the challenging invisible region. As soon as the ANNs involve the invisibility point in the span of parameters, the recognition abilities decrease significantly (Figure 6c,d) due to the challenges discussed above for ANN$_4$.

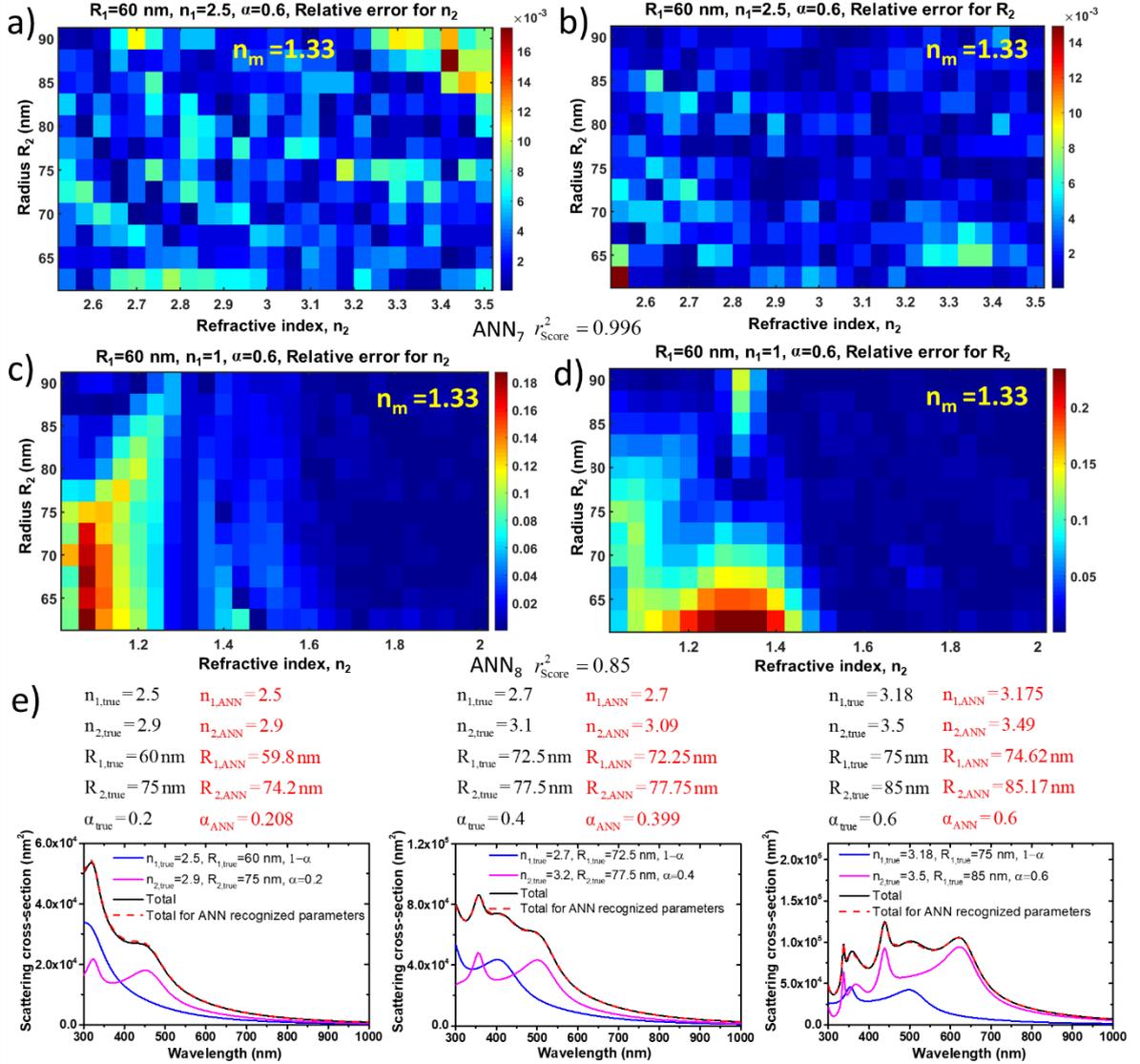

**Figure 6.** Five-parameter recognition scheme without and with the invisibility point. The figure depicts the relative-error maps for the refractive index ($n_2$) and the radius ($R_2$) obtained from two trained ANNs. In (a,b), the intervals of refractive indices are: $2.5 < n_1, n_2 < 3.5$. In panels (c,d), $1 < n_1, n_2 < 2$. Besides, for all panels, the radii intervals are $60\,\text{nm} < R_1, R_2 < 90\,\text{nm}$, and the concentration-coefficient interval is $0.2 < \alpha < 0.8$. For the matrix index, we take $n_m = 1.33$. (e) The examples of recognized and ground truth parameters of two different sizes NSs ($R_1$ and $R_2$) of different materials ($n_1$ and $n_2$), of $\alpha$ relative concentration. Four curves of scattering spectra show two spectra of NSs of $\alpha$ and $1-\alpha$ concentrations, the sum of these two spectra and the spectrum for the recognized parameters.

**2.3 Plasmonic Recognition**

A beam of light propagating through a medium containing metal nanostructures loses a fraction of its intensity due to the scattering and absorption induced by nano-objects. Therefore, colloidal solutions exhibit different colors, depending on the extinction profile and the plasmon resonance wavelength of nanoparticles.[49–51] The plasmonic resonances of metal NSs can be computed using the Mie theory.

*2.3.1 The basic two-parameter plasmonic scheme based on the Beer-Lambert law.*

The sensitivity of these resonances to the local refractive index led to the development of plasmonic sensors based on spectral shifts.[52–56] Within our proposed approach, simultaneously detecting the particle size, its density in the solution, and environmental RI, we can use a single measurement of an ensemble, which was not previously characterized optically, and extract information about the local dielectric environment. This can be useful in, e.g., devices operating under flow conditions where variant densities of nanoparticles' ensemble can traverse through. A careful analysis of such variation also requires tracking the widths and amplitudes of the resonances in the spectra, resulting in a non-trivial study of the data. Accordingly, we trained a neural network (ANN$_9$) for the recognition of the RI of the local environment of gold NSs ($R=60\,\mathrm{nm}$), simultaneously with the recognition of the NSs' density in solutions. In Figure 7a, we present the extinction spectra of gold NSs submerged in three various solutions, with refractive indexes of $n_\mathrm{m}=1.33$, $n_\mathrm{m}=1.5$ and $n_m = 1.7$. The corresponding densities and refractive indexes recognized by the neural network are shown on the right section of Figure 7a. It is important to note that the ANN$_9$ has a $r^2_{Score} = 0.9999$. This particularly high accuracy can be explained by the absence of the major challenges related to the invisibility point. In this case, the characteristic peak structure coming from the resonant plasmon excitations makes it easier to recognize the parameters. Whereas, for small and low-refractive index dielectric nanoparticles in the previous section (ANN$_1$-ANN$_4$), we dealt with spectra showing a monotonous decay behavior.

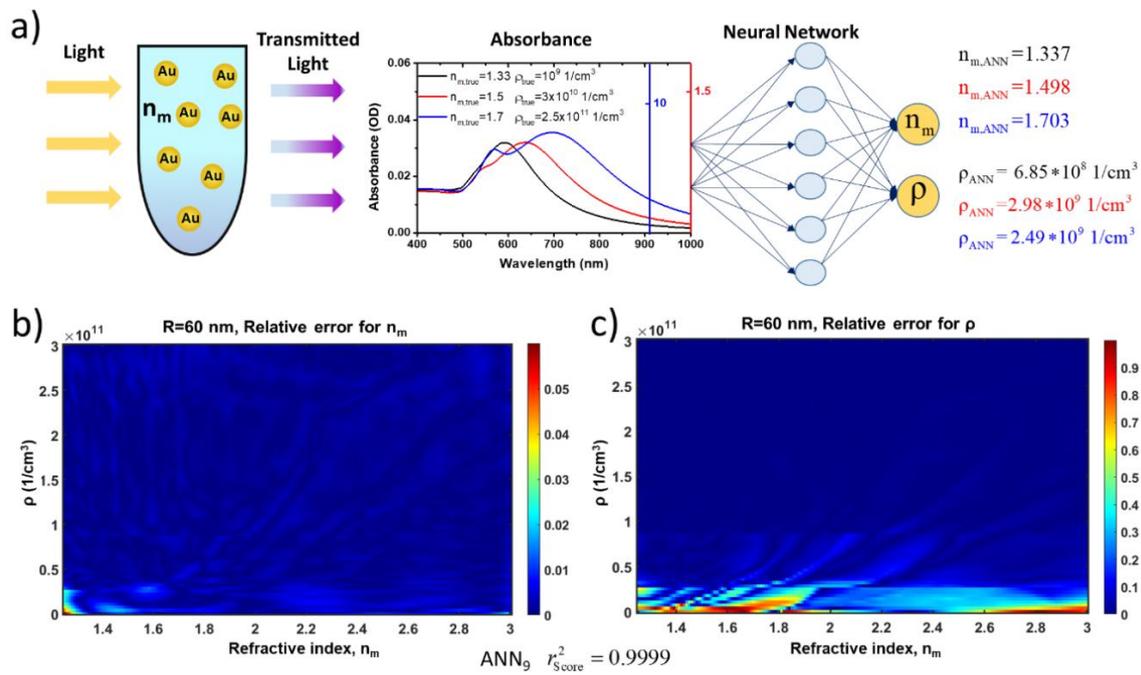

**Figure 7.** Two-parameter recognition scheme for the plasmonic sensing system, $(n_m, \rho)$. (a) Predicted refractive indexes $(n_m)$ of the matrix and the densities $(\rho)$ of the gold NS ($R=60$ nm) by the neural network from the extinction spectra. (b,c) Relative-error maps of the RI of the matrix and the map of the densities errors. In these graphs, we used the ANN$_9$ trained for the intervals $3*10^8 < \rho < 3*10^{11}$ and $1.25 < n_m < 3$. Optical path is set to be 1 cm.

Figure 7b, c shows the relative error maps of the predicted results for the refractive indices of the local environment and the densities of NSs. It is interesting to note a very small region with a noticeable error reflected on the map (Figure 7a). The maximum error of $n_m$-recognition is 0.066 for the smallest density of gold NSs submerged in the liquid of $n_m=1.25$. Although we observe errors for the recognition of $\rho$ for low densities (Figure 7b), the recognition of local environment ($n_m$) is accurate. Here we chose the range of local refractive indexes so that our model should reflect real experimental cases in solution and on a substrate. When the NSs are adsorbed on a substrate, the parameter $n_m$ should be regarded as an effective RI, averaged between substrate and medium. Thus, $n_m \approx 1.25$ should correspond to NSs on a glass substrate without any coverage. Higher refractive indices correspond to different solutions ($1.33 < n_m < 1.74$) and substrates with high RI covered by a solution. For instance, $n_m \approx 3$ can be assigned to a silicon substrate covered by a solution of interest. And variations from a given value will reflect changes in the local environment of the particles, such as analyte adsorption. Besides this 2-parameter recognition scheme $(n_m, \rho)$, one may see the results of the recognition of other 2 parameters, e.g., for RI of the local environment ($n_m$) of gold NSs and their sizes ($R$) (See **Figure S7**).

*2.3.2 The standard three-parameter plasmonic scheme.*

Next, we set up the case of plasmonic sensing considering a solution containing Au NSs of different sizes $(30\,\text{nm} < R < 100\,\text{nm})$ and diluted with different concentrations $(3*10^8\ 1/\text{cm}^3 < \rho < 3*10^{11}\ 1/\text{cm}^3)$. For this case, we restricted the refractive-index interval in the following way: $1.33 < n_m < 1.7$. The 2D maps show low (of the order of $10^{-3}$) and randomly scattered relative errors of the recognition for the matrix RI and the NS radii (**Figure 8a,b**). These 2D maps are retrieved by fixing one of the parameters; e.g., we take $\rho=2*10^{10}\,1/\text{cm}^3$ (see the cases for $\rho=2*10^9\,1/\text{cm}^3$ and $\rho=1.5*10^{11}\, 1/cm^3$ in **Figure S8**, the SI) from the three-dimensional data $(n_m, R, \rho)$ and obtain the relative errors for the remaining two parameters ($n_m$ and $\rho$). By analyzing the 2D error maps, the examples of the recognized parameters (Figure 8c), and the score ($r^2_{Score} = 0.999$), we may note that the ANN$_{10}$ successfully senses the solution containing different gold NSs of arbitrary concentrations.

One may find a four-parameter recognition scheme in **Figure S9** (SI), for the case of plasmonic sensing considering a solution containing Au NSs of two different sizes ($30\,nm < R_1, R_2 < 100\,nm$) and diluted with different relative concentrations ($\alpha$).

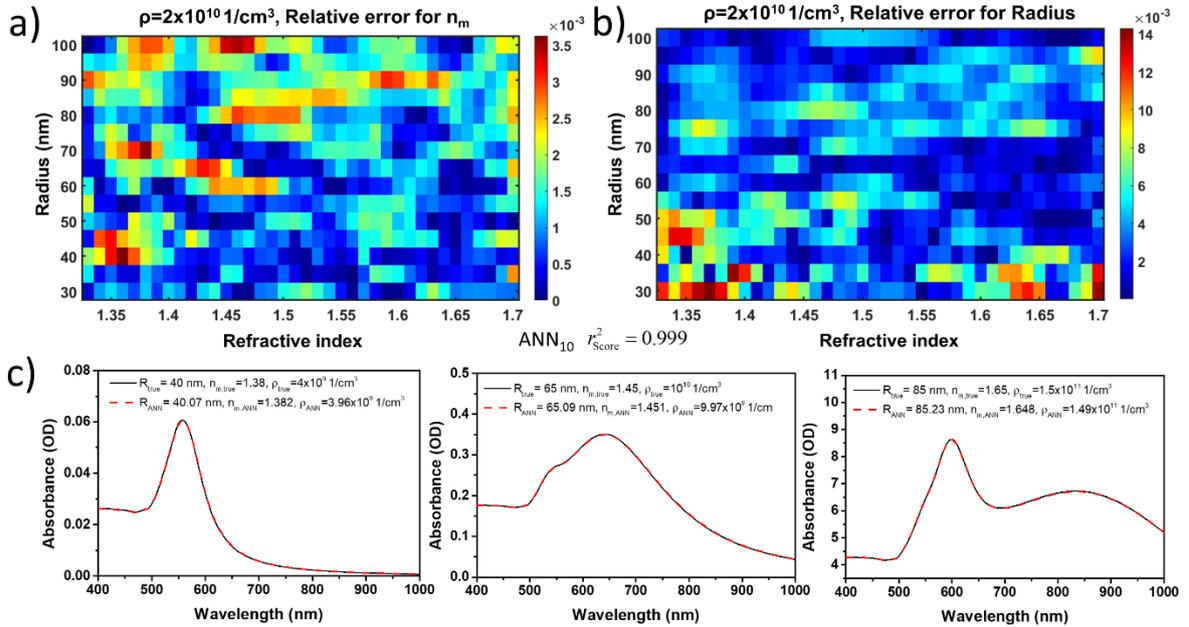

**Figure 8.** Three-parameter recognition model for the plasmonic sensing system, $(n_m, R, \rho)$. Relative-error maps for (a) the refractive-index of matrix ($n_m$) and (b) density of gold NSs ($\rho$) prediction obtained from the trained ANN$_{10}$ for the intervals of refractive indexes of matrix $1.33 < n_m < 1.7$, of radii $30\,nm < R < 100\,nm$, and of the concentration coefficient $3*10^8 < \rho < 3*10^{11}$. (c) The examples of recognized and ground truth parameters of gold NSs $(R, \rho,)$ and their local RI ($n_m$). Two curves show the absorbance spectrum for ground truth parameters (black solid line) and the spectrum plotted based on the recognized parameters by ANN$_{10}$ (red dashed-line).

## 3. Conclusions

Our approach using artificial neural networks showed success in the simultaneous recognition of two and more physical parameters such as the dimension of a nanosphere ($R$), the refractive index ($n$) of a matrix, the absolute ($\rho$) and relative ($\alpha$) concentrations of several species in a mixed ensemble. In particular, our extended ANN models recognized with high accuracy the mixture of two dissimilar NSs and their concentration in an aqueous solution from an optical spectrum, which corresponds to the recognition of five parameters. The extension of the number of recognized parameters leads to the restriction of the range of the employing intervals of radii ($R_{max} - R_{min} = 30\,\text{nm}$) and RI ($n_{max} - n_{min} = 1$) due to the progressive augmentation of the input spectra amount with a single parameter addition.

This work enables the utilizing artificial neural networks in a new direction, for example, for the rapid, real-time recognition of nano-objects for atmospheric and marine analysis. Another possible use of our predictions concerns plasmonic bio-sensing and molecular analysis. The integration of a well-trained ANN into a spectral detection system can pave the way towards new autonomous, low-power sensing devices for fast and accurate recognition of matter and nano-objects from a single optical spectrum. Our results examine limiting cases where the parameters of nano-objects are close to the invisibility point regime and beyond. ANNs operating outside the invisibility-point regime showed systematically high scores. We also showed that the ANNs can recognize NSs in air nearby the invisibility point conditions with a high score. Fundamental limitations were found and described for the recognition of the NSs submerged in water in the following regimes: (1) the case when the optical spectra of NSs are non-unique and (2) the physical situation when the refractive index of the NSs is close to that of the matrix (the invisibility point). And as a result, we have formulated the main challenges in training this kind of system. However, despite the above limitations, the trained ANN showed overall a high recognition accuracy with a high score. To give more details, one of the major challenges in training this kind of system comes from the following factor. Some of the NSs do not show resonances in the chosen optical range ($300\,\text{nm} < \lambda < 1000\,\text{nm}$) and, therefore, have non-characteristic, low-amplitude, and featureless spectra, which are difficult to recognize. Along with dielectric nano-objects, we demonstrated the ability of ANNs for the recognition of the dielectric constant of the matrix using embedded metal nanospheres and their mixtures; in this plasmonic scheme, the recognition process reproduces both the nanosphere sizes and their concentrations. Overall, the ANNs have showed an excellent ability to predict the required physical parameters. Finally, we think that the ML-based schemes described in this study can be employed in various sensing applications (plasmonic and others), enhancing their accuracy.

References

[1]  G. Mie, *Ann. Phys.* **1908**, *330*, 377.

[2]  W. Hergert, T. Wriedt, *The Mie Theory: Basics and Applications*, Springer, **2012**.

[3]  W. Chaâbani, J. Proust, A. Movsesyan, J. Béal, A.-L. Baudrion, P.-M. Adam, A. Chehaidar, J. Plain, *ACS Nano* **2019**, *13*, 4199.

[4]  S. A. Maier, *Plasmonics: Fundamentals and Applications*, Springer, New York, **2007**.

[5]  P. Mulvaney, G. Hartland, *Phys. Chem. Chem. Phys.* **2009**, *11*, 5866.

[6]  A. Movsesyan, A. Muravitskaya, M. Castilla, S. Kostcheev, J. Proust, J. Plain, A. L. Baudrion, R. Vincent, P. M. Adam, *J. Phys. Chem. C* **2021**, *125*, 724.

[7]  D. Tzarouchis, A. Sihvola, *Appl. Sci.* **2018**, *8*, DOI: 10.3390/app8020184.

[8]  A. Al-Zubeidi, L. A. McCarthy, A. Rafiei-Miandashti, T. S. Heiderscheit, S. Link, *Single-Particle Scattering Spectroscopy: Fundamentals and Applications*, **2021**.

[9]  P. C. Waterman, *Phys. Rev. D* **1971**, *3*, 825.

[10] K. S. Yee, *IEEE Trans. Antennas Propag.* **1966**, *14*, 302.

[11] D. Liu, L. H. Gabrielli, M. Lipson, S. G. Johnson, *Opt. Express* **2013**, *21*, 14223.

[12] A. Y. Piggott, J. Lu, K. G. Lagoudakis, J. Petykiewicz, T. M. Babinec, J. Vucković, *Nat. Photonics* **2015**, *9*, 374.

[13] A. Y. Piggott, J. Petykiewicz, L. Su, J. Vučković, *Sci. Rep.* **2017**, *7*, 1.

[14] J. Lu, J. Vučković, *Opt. Express* **2013**, *21*, 13351.

[15] S. Molesky, Z. Lin, A. Y. Piggott, W. Jin, J. Vucković, A. W. Rodriguez, *Nat. Photonics* **2018**, *12*, 659.

[16] C. M. Lalau-Keraly, S. Bhargava, O. D. Miller, E. Yablonovitch, *Opt. Express* **2013**, *21*, 21693.

[17] A. Mehrabian, M. Miscuglio, Y. Alkabani, V. J. Sorger, T. El-Ghazawi, *IEEE J. Sel. Top. Quantum Electron.* **2020**, *26*, DOI: 10.1109/JSTQE.2019.2957443.

[18] S. So, T. Badloe, J. Noh, J. Rho, J. Rho, J. Bravo-Abad, *Nanophotonics* **2020**, *9*, 1041.

[19] P. R. Wiecha, A. Lecestre, N. Mallet, G. Larrieu, *Nat. Nanotechnol.* **2019**, *14*, 237.

[20] J. Schmidt, M. R. G. Marques, S. Botti, M. A. L. Marques, *npj Comput. Mater.* **2019**, *5*, DOI: 10.1038/s41524-019-0221-0.

[21] B. Sanchez-Lengeling, A. Aspuru-Guzik, *Science (80-. ).* **2018**, *361*, 360.

[22] A. Tittl, A. John-Herpin, A. Leitis, E. R. Arvelo, H. Altug, *Angew. Chemie - Int. Ed.* **2019**, *58*, 14810.

[23] A. Turpin, I. Vishniakou, J. d. Seelig, *Opt. Express* **2018**, *26*, 30911.

[24] C. S. Ho, N. Jean, C. A. Hogan, L. Blackmon, S. S. Jeffrey, M. Holodniy, N.

Banaei, A. A. E. Saleh, S. Ermon, J. Dionne, *Nat. Commun.* **2019**, *10*, DOI: 10.1038/s41467-019-12898-9.

[25] J. Peurifoy, Y. Shen, L. Jing, Y. Yang, F. Cano-Renteria, B. G. DeLacy, J. D. Joannopoulos, M. Tegmark, M. Soljačić, *Sci. Adv.* **2018**, *4*, 1.

[26] T. Zhang, J. Wang, Q. Liu, J. Zhou, J. Dai, X. Han, Y. Zhou, K. Xu, *arXiv* **2018**, *7*, 368.

[27] E. Ashalley, K. Acheampong, L. V. Besteiro, P. Yu, A. Neogi, A. O. Govorov, Z. Wang, *Photonics Res.* **2020**, *8*, 1213.

[28] W. Ma, F. Cheng, Y. Liu, *ACS Nano* **2018**, *12*, 6326.

[29] S. So, J. Mun, J. Rho, *ACS Appl. Mater. Interfaces* **2019**, *11*, 24264.

[30] J. Timoshenko, D. Lu, Y. Lin, A. I. Frenkel, *J. Phys. Chem. Lett.* **2017**, *8*, 5091.

[31] Z. A. Kudyshev, S. Bogdanov, T. Isacsson, A. Kildishev, A. Boltasseva, V. M. Shalaev, *Adv. Quantum Technol.* **2020**, *3*.

[32] C. L. Chen, A. Mahjoubfar, L. C. Tai, I. K. Blaby, A. Huang, K. R. Niazi, B. Jalali, *Sci. Rep.* **2016**, *6*, 1.

[33] W. Ma, F. Cheng, Y. Xu, Q. Wen, Y. Liu, *Adv. Mater.* **2019**, *31*, 1.

[34] Z. Liu, D. Zhu, S. P. Rodrigues, K. T. Lee, W. Cai, *Nano Lett.* **2018**, *18*, 6570.

[35] L. V. Besteiro, P. Yu, Z. Wang, A. W. Holleitner, G. V. Hartland, G. P. Wiederrecht, A. O. Govorov, *Nano Today* **2019**, *27*, 120.

[36] A. O. Govorov, H. H. Richardson, *Nano Today* **2007**, *2*, 30.

[37] Y. Lecun, Y. Bengio, G. Hinton, *Nature* **2015**, *521*, 436.

[38] Y. Zhang, J. Gao, H. Zhou, *Adv. Neural Inf. Process. Syst.* **2012**, *25*, 1097.

[39] C. Couprie, L. Najman, Y. Lecun, *Pattern Anal. Mach. Intell. IEEE Trans.* **2013**, *35*, 1915.

[40] W. Ouyang, A. Aristov, M. Lelek, X. Hao, C. Zimmer, *Nat. Biotechnol.* **2018**, *36*, 460.

[41] H. Wang, Y. Rivenson, Y. Jin, Z. Wei, R. Gao, H. Günaydın, L. A. Bentolila, C. Kural, A. Ozcan, *Nat. Methods* **2019**, *16*, 103.

[42] T. Asano, S. Noda, *arXiv* **2018**, *26*, 32704.

[43] D. Liu, Y. Tan, E. Khoram, Z. Yu, *ACS Photonics* **2018**, *5*, 1365.

[44] C. F. Bohren, D. R. Huffman, *Absorption and Scattering of Light by Small Particles*, Wiley, New York, **1983**.

[45] D. K. Barupal, O. Fiehn, *J. Mach. Learn. Res.* **2011**, *12*, 2825.

[46] J. Baxter, A. Calà Lesina, J. M. Guay, A. Weck, P. Berini, L. Ramunno, *Sci. Rep.* **2019**, *9*, 1.

[47] I. Malkiel, M. Mrejen, A. Nagler, U. Arieli, L. Wolf, H. Suchowski, *Light Sci. Appl.* **2018**, *7*, DOI: 10.1038/s41377-018-0060-7.

[48] O. Hemmatyar, S. Abdollahramezani, Y. Kiarashinejad, M. Zandehshahvar, A. Adibi, *Nanoscale* **2019**, *11*, 21266.

[49] S. E. Skrabalak, L. Au, X. Li, Y. Xia, *Nat. Protoc.* **2007**, *2*, 2182.

[50] R. S. Moirangthem, M. T. Yaseen, P.-K. Wei, J.-Y. Cheng, Y.-C. Chang, *Biomed. Opt. Express* **2012**, *3*, 899.

[51] X. Huang, M. A. El-Sayed, *J. Adv. Res.* **2010**, *1*, 13.

[52] J. N. Anker, W. P. Hall, O. Lyandres, N. C. Shah, J. Zhao, R. P. Van Duyne, *Nat. Mater.* **2008**, *7*, 442.

[53] K. M. Mayer, J. H. Hafner, *Sensors Actuators B Chem.* **1999**, *54*, 3.

[54] J. Jatschka, A. Dathe, A. Csáki, W. Fritzsche, O. Stranik, *Sens. Bio-Sensing Res.* **2016**, *7*, 62.

[55] J. Otsuki, K. Sugawa, S. Jin, *Mater. Adv.* **2021**, *2*, 32.

[56] A. V. Kabashin, P. Evans, S. Pastkovsky, W. Hendren, G. A. Wurtz, R. Atkinson, R. Pollard, V. A. Podolskiy, A. V. Zayats, *Nat. Mater.* **2009**, *8*, 867.

[57] P. B. Johnson, R. W. Christy, *Phys. Rev. B* **1972**, *6*, 4370.

# Supporting Information

**Mie Sensing with Neural Networks: Recognition of Nano-Object Parameters, the Invisibility Point, and Restricted Models**

*Artur Movsesyan, Lucas V. Besteiro, Zhiming Wang,* and Alexander O. Govorov**

**Methods: Database preparation using the Mie theory.**

The datasets used for the training and validation of our ANNs were composed by synthetic spectra, created with the Mie Theory. This formalism, which presents an exact solution of the Maxwell equations for spheres by considering the multipole expansion of the incident planewave and the scattered field into vector spherical harmonics, provides an analytical method for the calculation of the optical response of spherical nano-objects.[2,44] This approach allows us to prepare a large number of spectra in a very short time. For instance, a modern consumer computer can easily calculate hundreds of such spectra per second. In the first part of our study, we considered nanoparticles without an imaginary part in their dielectric function. Therefore, the optical response of dielectric NSs contains only scattering. The scattering cross-section is given by the following expression, within the Mie theory formalism (see section 4.4 in Ref. [44]):

$$\sigma_{\text{Sca}} = \frac{2\pi}{k_0^2} \sum_{j=1}^{n} (2j+1)\left(\left|a_j\right|^2 + \left|b_j\right|^2\right), \quad \text{(S1)}$$

where $k_0$ is the wave vector, $a_j$ and $b_j$ are the electric and magnetic Mie coefficients of the $j$-th multipolar order. The Mie coefficients are given by the following equations:[44]

$$a_j = \frac{m\psi_j(k_0,R,m)\psi_j^{'}(k_0,R) - \psi_j(k_0,R)\psi_j^{'}(k_0,R,m)}{m\psi_j(k_0,R,m)\xi_j^{'}(k_0,R) - \xi_j(k_0,R)\psi_j^{'}(k_0,R,m)},$$

$$b_j = \frac{\psi_j(k_0, R, m)\psi_j'(k_0, R) - m\psi_j(k_0, R)\psi_j'(k_0, R, m)}{\psi_j(k_0, R, m)\xi_j'(k_0, R) - m\xi_j(k_0, R)\psi_j'(k_0, R, m)}, \quad (S2)$$

where $\psi_j$ and $\xi_j$ are the Ricatti–Bessel functions, $R$ is the radius of the NS, and $m$ is the ratio between the refractive indices of the NS ($n$) and the surrounding environment ($n_m$).

The optical response of metal nanostructures due to their lossy nature is characterized by their extinction, which is the sum of the absorption and the scattering. The extinction spectrum of a NS has the following form:[44]

$$\sigma_{Ext} = \frac{2\pi}{k_0^2} \sum_{j=1}^{n} (2j+1) \text{Re}(a_j + b_j),$$

where $k_0$ is the wave vector of the impinging light, $a_j$ and $b_j$ are the electric and magnetic Mie coefficients of the j-th multipolar order.

We have prepared 4 distinct datasets for the ANN training (ANN$_1$-ANN$_4$) for two parameter recognition schemes; the first two datasets (ANN$_1$, ANN$_3$) describe dielectric nanospheres embedded in air and the remaining two (ANN$_2$, ANN$_4$) consider water as the medium. In both cases, we fixed an interval of refractive indices for the nano-objects. The nanospheres' radii were taken in the interval from 30nm to 100nm, paced by 2 nm, for these four datasets. The spectral range of the optical scattering spectra is fixed between 300nm and 1000nm.

Then, we prepared four datasets for the ANN training (ANN$_5$, ANN$_6$, ANN$_{11}$, ANN$_{12}$) for three parameter recognition schemes respectively for the following intervals of the refractive indexes $2.5 < n < 4$ (ANN$_5$, ANN$_{11}$) and $1 < n < 2.5$ (ANN$_6$, ANN$_{12}$). The concentration span was fixed $4*10^8\,1/\text{cm}^3 < \rho < 4*10^{11}\,1/\text{cm}^3$ (ANN$_5$) for the first interval of refractive indexes ($2.5 < n < 4$), and for the latter interval ($1 < n < 2.5$) it was $2*10^9\,1/\text{cm}^3 < \rho < 4*10^{12}\,1/\text{cm}^3$ (ANN$_6$). The nanospheres' radii were taken in the interval from 50nm to 100nm $(50\,\text{nm} < R < 100\,\text{nm})$, paced by 2.5 nm. ANN$_{11}$, ANN$_{12}$ were trained to recognize the mixture of two dielectric nanospheres of different sizes of similar concentration and their material. For this study, two nanospheres' radii were taken in the interval $50\,\text{nm} < R_1, R_2 < 100\,\text{nm}$.

The two datasets were prepared for the ANN training (ANN$_7$, ANN$_8$) of the recognition scheme of five parameters. The intervals of the parameters were the following for the ANN$_7$ and ANN$_8$: $[2.5 < n_1, n_2 < 3.5]$ $60\,\text{nm} < R_1, R_2 < 90\,\text{nm}$, $0.2 < \alpha < 0.8$. $\alpha$ is the concentration coefficient of the NS$_2$ ($R_2, n_2$). Hence, the concentration of NS$_1$ ($R_1, n_1$) is $1-\alpha$. The intervals for RI were set $2.5 < n_1, n_2 < 3.5$ and $1 < n_1, n_2 < 2$ for ANN$_7$ and ANN$_8$, respectively.

We have also made a dataset for the training of the ANNs (ANN$_9$, ANN$_{10}$, ANN$_{13}$, ANN$_{14}$) considering gold nanospheres embedded in the media with refractive indices respectively within in the intervals $1.25 < n_m < 3$ (ANN$_9$, ANN$_{13}$) and $1.33 < n_m < 1.7$ (ANN$_{10}$, ANN$_{14}$). For these cases of ANN$_{10}$, ANN$_{13}$, ANN$_{14}$ the radii of gold nanospheres were taken in and interval $30\,\text{nm} < R_1, R_2 < 100\,\text{nm}$ and for ANN$_9$ respectively $15\,\text{nm} < R < 100\,\text{nm}$. In our computations, we used experimental tables for the complex permittivity of gold.[57] One may see all details of datasets for the trainings presented in the table below.

**Table S2**. The configuration of the datasets for the ANN training, validation and the scores for testing sets.

| ANN | $R$-interval | $n$-interval | $\alpha$-interval $\rho$-interval | Matrix RI | Total Spectra | $r^2_{Score}$ |
|---|---|---|---|---|---|---|
| ANN$_1$ | $30\,\text{nm} < R < 100\,\text{nm}$ | $1.35 < n < 5$ | n/a | $n_m = 1$ | 13212 | 0.9993 |
| ANN$_2$ | $30\,\text{nm} < R < 100\,\text{nm}$ | $1.5 < n < 5$ | n/a | $n_m = 1.33$ | 12636 | 0.9985 |
| ANN$_3$ | $30\,\text{nm} < R < 100\,\text{nm}$ | $1 < n < 5$ | n/a | $n_m = 1$ | 14436 | 0.9977 |
| ANN$_4$ | $30\,\text{nm} < R < 100\,\text{nm}$ | $1 < n < 5$ | n/a | $n_m = 1.33$ | 14436 | 0.9854 |
| ANN$_5$ | $50\,\text{nm} < R < 100\,\text{nm}$ | $2.5 < n < 4$ | $4*10^8\,1/\text{cm}^3 < \rho < 4*10^{11}\,1/\text{cm}^3$ | $n_m = 1.33$ | 63840 | 0.9994 |
| ANN$_6$ | $50\,\text{nm} < R < 100\,\text{nm}$ | $1 < n < 2.5$ | $2*10^9\,1/\text{cm}^3 < \rho < 4*10^{12}\,1/\text{cm}^3$ | $n_m = 1.33$ | 63840 | 0.867 |
| ANN$_7$ | $60\,\text{nm} < R_1, R_2 < 90\,\text{nm}$ | $2.5 < n_1, n_2 < 3.5$ | $0.2 < \alpha < 0.8$ | $n_m = 1.33$ | 101400 | 0.996 |
| ANN$_8$ | $60\,\text{nm} < R_1, R_2 < 90\,\text{nm}$ | $1 < n_1, n_2 < 2$ | $0.2 < \alpha < 0.8$ | $n_m = 1.33$ | 101400 | 0.85 |
| ANN$_9$ | $R = 60\,\text{nm}$ | $n = n_{Au}$ | $3*10^8\,1/\text{cm}^3 < \rho < 3*10^{11}\,1/\text{cm}^3$ | $1.25 < n_m < 3$ | 15136 | 0.9999 |
| ANN$_{10}$ | $30\,\text{nm} < R < 100\,\text{nm}$ | $n = n_{Au}$ | $3*10^8\,1/\text{cm}^3 < \rho < 3*10^{11}\,1/\text{cm}^3$ | $1.33 < n_m < 1.7$ | 49020 | 0.99 |
| ANN$_{11}$ | $50\,\text{nm} < R_1, R_2 < 100\,\text{nm}$ | $2.5 < n < 4$ | n/a | $n_m = 1.33$ | 24700 | 0.9997 |
| ANN$_{12}$ | $50\,\text{nm} < R_1, R_2 < 100\,\text{nm}$ | $1 < n < 2.5$ | n/a | $n_m = 1.33$ | 24700 | 0.911 |
| ANN$_{13}$ | $15\,\text{nm} < R < 100\,\text{nm}$ | $n = n_{Au}$ | n/a | $1.25 < n_m < 3$ | 12285 | 0.9999 |
| ANN$_{14}$ | $30\,\text{nm} < R_1, R_2 < 100\,\text{nm}$ | $n = n_{Au}$ | $0.1 < \alpha < 0.9$ | $1.33 < n_m < 1.7$ | 35910 | 0.99 |

The optical spectra of the training datasets were normalized by the maximum value of the optical cross-section ($0 < \sigma_{\text{Sca\_norm}} < 1$). To exclude the effect of different scales of the output parameters (radius and refractive index), we applied the well-established z-score standardization scheme to treat the output data.

$$y_{\text{standard}} = \frac{y - \bar{y}}{\sigma},$$

where $y$ describes the original data, $\bar{y}$ is the mean value of the population, and $\sigma$ is the standard deviation of the population.

**ANN architecture.**

The chosen ANN for the two parameters (radius and refractive index) recognition scheme is a deep neural network consisting of the input and output layers, and 4 hidden layers. We have trained ANNs, for the following cases: ANN$_1$ ($n_m=1$, $30\,\text{nm} < R < 100\,\text{nm}$, $1.35 < n < 5$), ANN$_2$ ($n_m = 1.33$, $30\,\text{nm} < R < 100\,\text{nm}$, $1.5 < n < 5$), ANN$_3$ ($n_m = 1$, $30\,\text{nm} < R < 100\,\text{nm}$, $1 < n < 5$), ANN$_4$ ($n_m = 1.33$, $30\,\text{nm} < R < 100\,\text{nm}$, $1.5 < n < 5$) ANN$_9$ ($1.25 < n_m < 3$, $R=60\,\text{nm}$, $3*10^8\,1/\text{cm}^3 < \rho < 3*10^{11}\,1/\text{cm}^3$, $n = n_{\text{Au}}$) and ANN$_{13}$ ($1.25 < n_m < 3$, $15\,\text{nm} < R < 100\,\text{nm}$, $n = n_{\text{Au}}$).

The configuration of the hidden layers of these five ANNs (ANN$_1$-ANN$_4$, ANN$_9$) have the following form: 200-500-200-30. We used "Adam" as a solver. The learning rate was fixed at 0.0005. For the remaining ANNs: ANN$_5$ ($n_m = 1.33$, $2.5 < n < 4$, $50\,\text{nm} < R < 100\,\text{nm}$, $3*10^8\,1/\text{cm}^3 < \rho < 3*10^{11}\,1/\text{cm}^3$), ANN$_6$ ($n_m = 1.33$, $1 < n < 2.5$, $50\,\text{nm} < R < 100\,\text{nm}$, $3*10^8\,1/\text{cm}^3 < \rho < 3*10^{11}\,1/\text{cm}^3$), ANN$_7$ ($n_m = 1.33$, $60\,\text{nm} < R_1, R_2 < 90\,\text{nm}$, $2.5 < n_1, n_2 < 3.5$, $0.2 < \alpha < 0.8$), ANN$_8$ ($n_m=1.33$, $60\,\text{nm} < R_1, R_2 < 90\,\text{nm}$, $1 < n_1, n_2 < 2$, $0.2 < \alpha < 0.8$), ANN$_{10}$ ($1.33 < n_m < 1.7$, $3*10^8\,1/\text{cm}^3 < \rho < 3*10^{11}\,1/\text{cm}^3$, $30\,\text{nm} < R < 100\,\text{nm}$, $n = n_{\text{Au}}$), ANN$_{11}$ ($n_m = 1.33$, $2.5 < n < 4$, $50\,\text{nm} < R_1, R_2 < 100\,\text{nm}$), ANN$_{12}$ ($n_m = 1.33$, $50\,\text{nm} < R_1, R_2 < 100\,\text{nm}$, $1 < n < 2.5$), ANN$_{14}$ ($1.33 < n_m < 1.7$, $30\,\text{nm} < R_1, R_2 < 100\,\text{nm}$, $0.1 < \alpha < 0.9$, $n = n_{Au}$), it was constructed five hidden layers with the following neural configuration 200-400-400-200-30.

We used 75% of the datasets (total Mie spectra) for the training and validation, and the remaining 25% for the final testing. The datasets were split randomly. 90% from this 75% was used for the training of ANNs. The 10% from 75% was used for the validation to avoid the overfitting, and an early termination of the training was applied after 50 consecutive non-improving score registration based on the validation score.

The accuracy of the ANN was determined by the coefficient of the determination ($r^2_{Score}$) using only the dataset of the testing (25% of total Mie spectra).

$$r^2_{Score} = 1 - \frac{\sum_i (y^2_{true_i} - y^2_{pred_i})}{\sum_i (y^2_{true_i} - \bar{y}^2_{pred})},$$

where $y_{true_i}$ is the expected value of $R$ and $n$, $y_{pred_i}$ is the predicted value, and $\bar{y}_{pred}$ is the mean value of the predicted results.

**Table S2.** The configuration and the scores of different ANNs for $n_m = 1$, $30\,\text{nm} < R < 100\,\text{nm}$, and $1.35 < n < 5$.

| ANNs | Configuration of hidden layers | $r^2_{Score}$ |
|---|---|---|
| ANN$_1$ (used in main manuscript) | 200-500-200-30 | 0.9993 |
| ANN$_i$ (more neurons) | 400-1000-400-60 | 0.9972 |
| ANN$_{ii}$ (less neurons) | 100-250-100-15 | 0.9987 |
| ANN$_{iii}$ (more layers - 6) | 200-400-500-400-200-30 | 0.9984 |
| ANN$_{iv}$ (less layers - 3) | 200-500-30 | 0.9983 |

| ANN$_v$ (less layers - 2) | 200-30 | 0.9967 |

The mean amplitude of scattering spectra is calculated by the following formalism:

$$\sigma_{Mean} = \frac{\sum_i \sigma_i}{N},$$

where $\sigma_i$ is the scattering cross-section of the $i$-th wavelength, and $N$ is the total number of wavelength points in a spectrum.

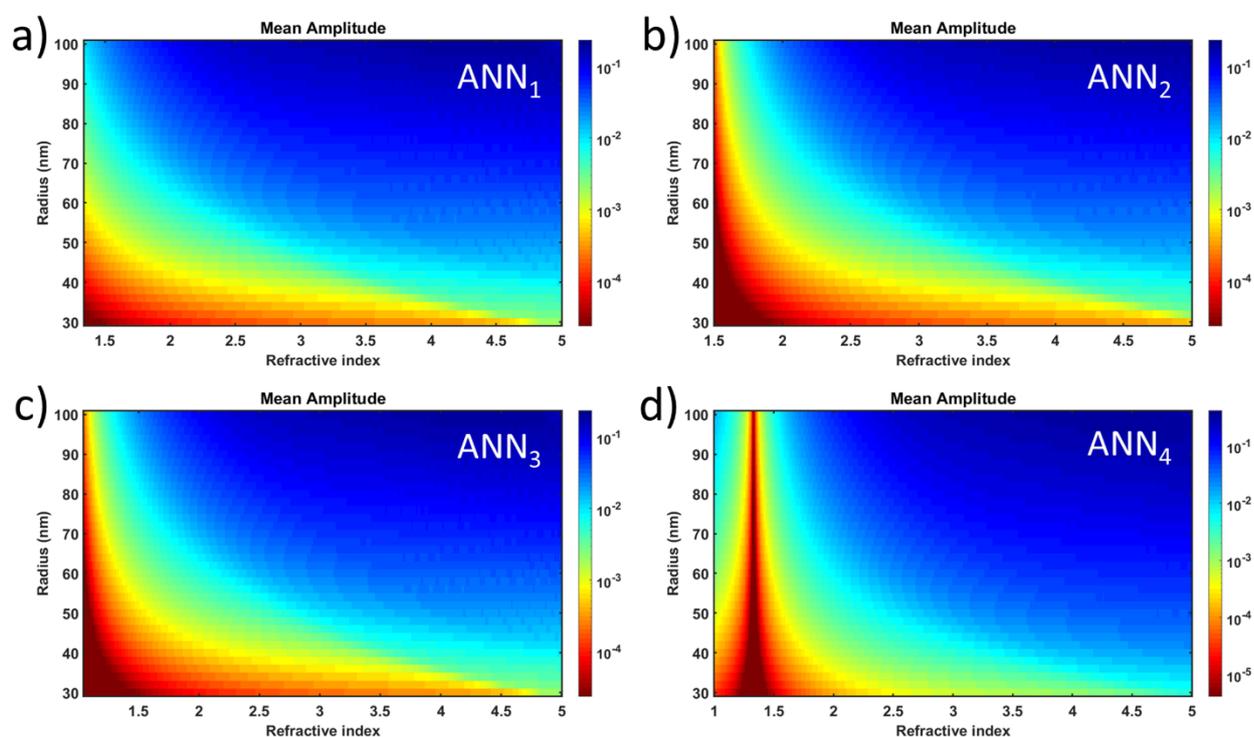

**Figure S1.** Mean amplitudes of normalized scattering of dielectric nanospheres used for the training of ANN$_1$-ANN$_4$.

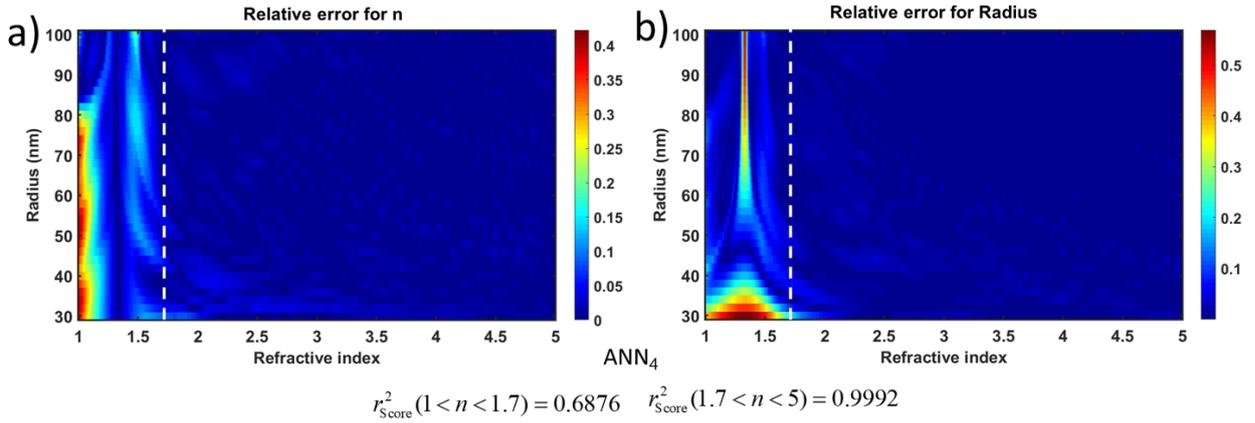

$r^2_{\text{Score}}(1<n<1.7) = 0.6876 \quad r^2_{\text{Score}}(1.7<n<5) = 0.9992$

**Figure S2.** Relative-error map of recognized results by ANN$_4$. $r^2_{\text{Score}}$ is evaluated in the two intervals $1<n<1.7$ and $1.7<n<5$. The challenging regime (near the transmission window) shows a comparably poor score of validation ($r^2_{\text{Score}} = 0.6876$), while out of this regime the score is high ($r^2_{\text{Score}} = 0.9992$). White dashed line shows the separation point ($n = 1.7$) of the two regimes.

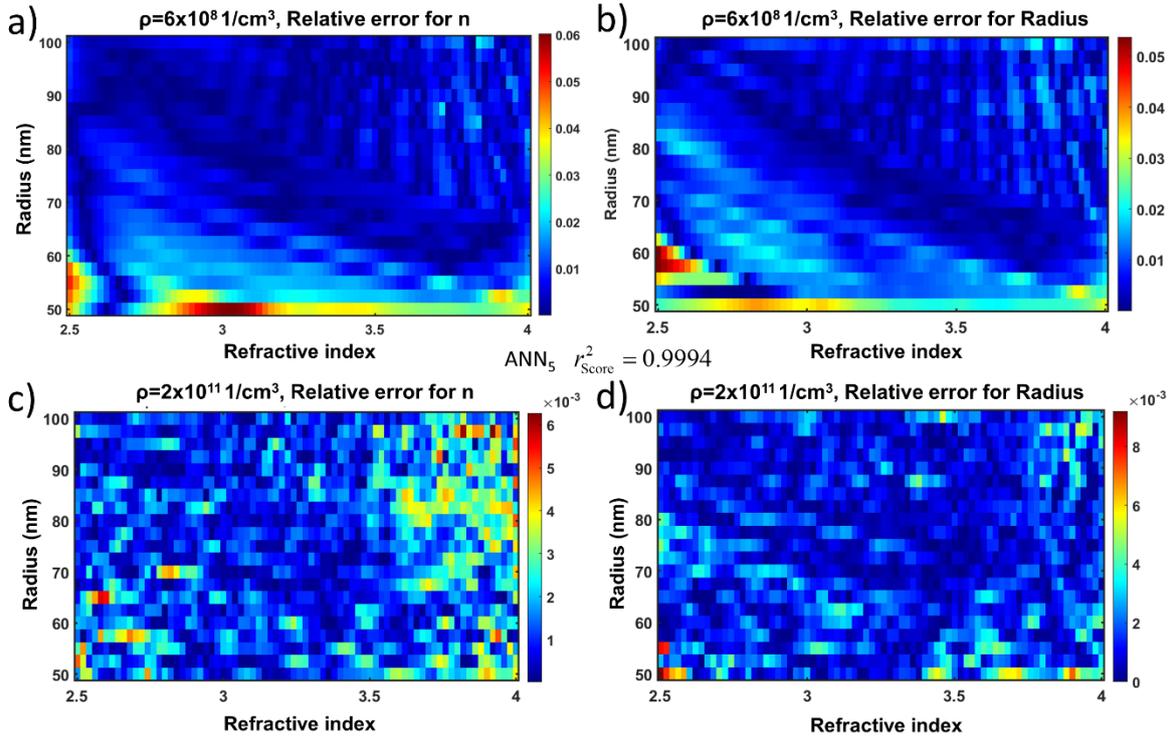

**Figure S3.** Three-parameter recognition scheme for NSs $(R, n, \rho)$. The figure shows the relative-error maps for the refractive-index ($n$) and radius prediction ($R$) obtained from the trained ANN (ANN$_5$) for the interval of refractive indexes: $2.5<n<4$; the radii interval is $50\,\text{nm}<R<100\,\text{nm}$. The density interval is $4*10^8\,1/\text{cm}^3 < \rho < 4*10^{11}\,1/\text{cm}^3$. For the matrix RI, $n_m = 1.33$. To make these 2D maps, we fixed one of the parameters, e.g., the density ($\rho$), since we deal with three-dimensional data. For panel (a,b) $\rho = 6*10^8\,1/\text{cm}^3$ and for (c,d) $\rho = 2*10^{11}\,1/\text{cm}^3$.

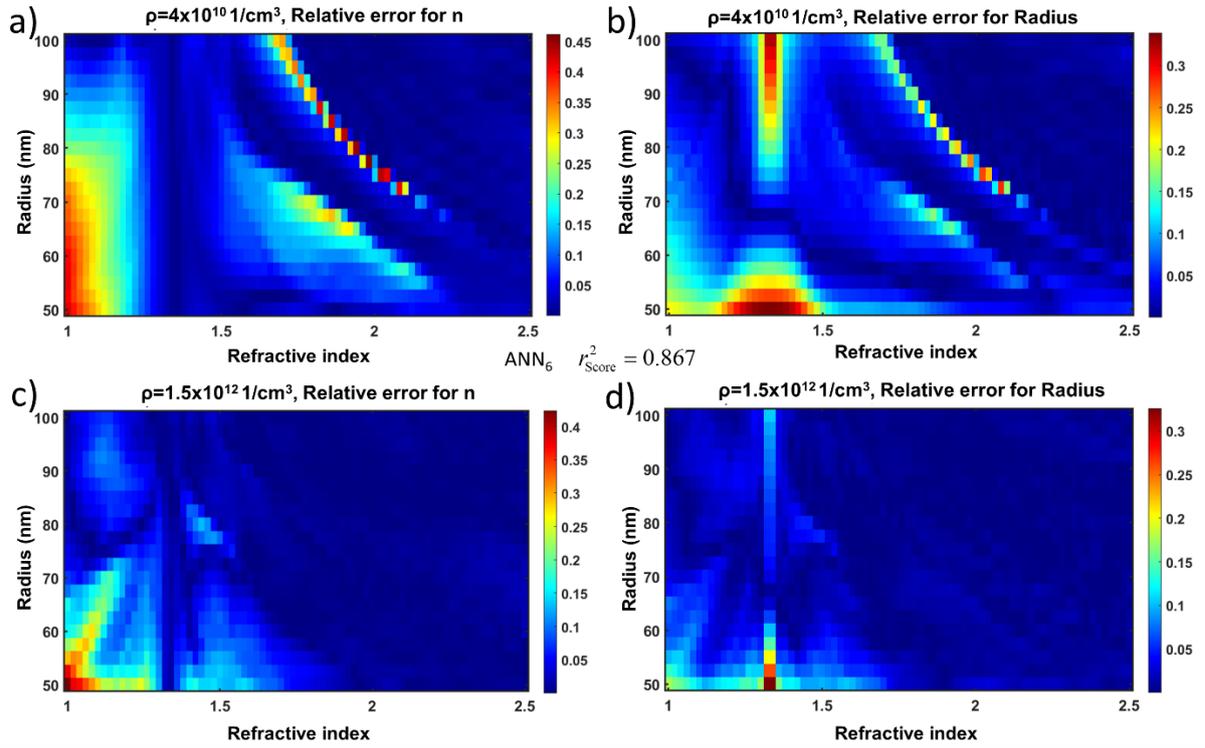

**Figure S4.** Three-parameter recognition scheme for NSs ($R, n, \rho$). The figure shows the relative-error maps for the refractive-index ($n$) and radius prediction ($R$) obtained from the trained ANN (ANN$_6$) for the interval of refractive indexes: $1 < n < 2.5$; the radii interval is $50\,\text{nm} < R < 100\,\text{nm}$. The density interval is $2*10^9\,1/\text{cm}^3 < \rho < 4*10^{12}\,1/\text{cm}^3$. For the matrix, $n_m = 1.33$. To make these 2D maps, we fixed one of the parameters, e.g., the density ($\rho$). For panel (a,b) $\rho = 4*10^{10}\,1/\text{cm}^3$ and for (c,d) $\rho = 4*10^{12}\,1/\text{cm}^3$.

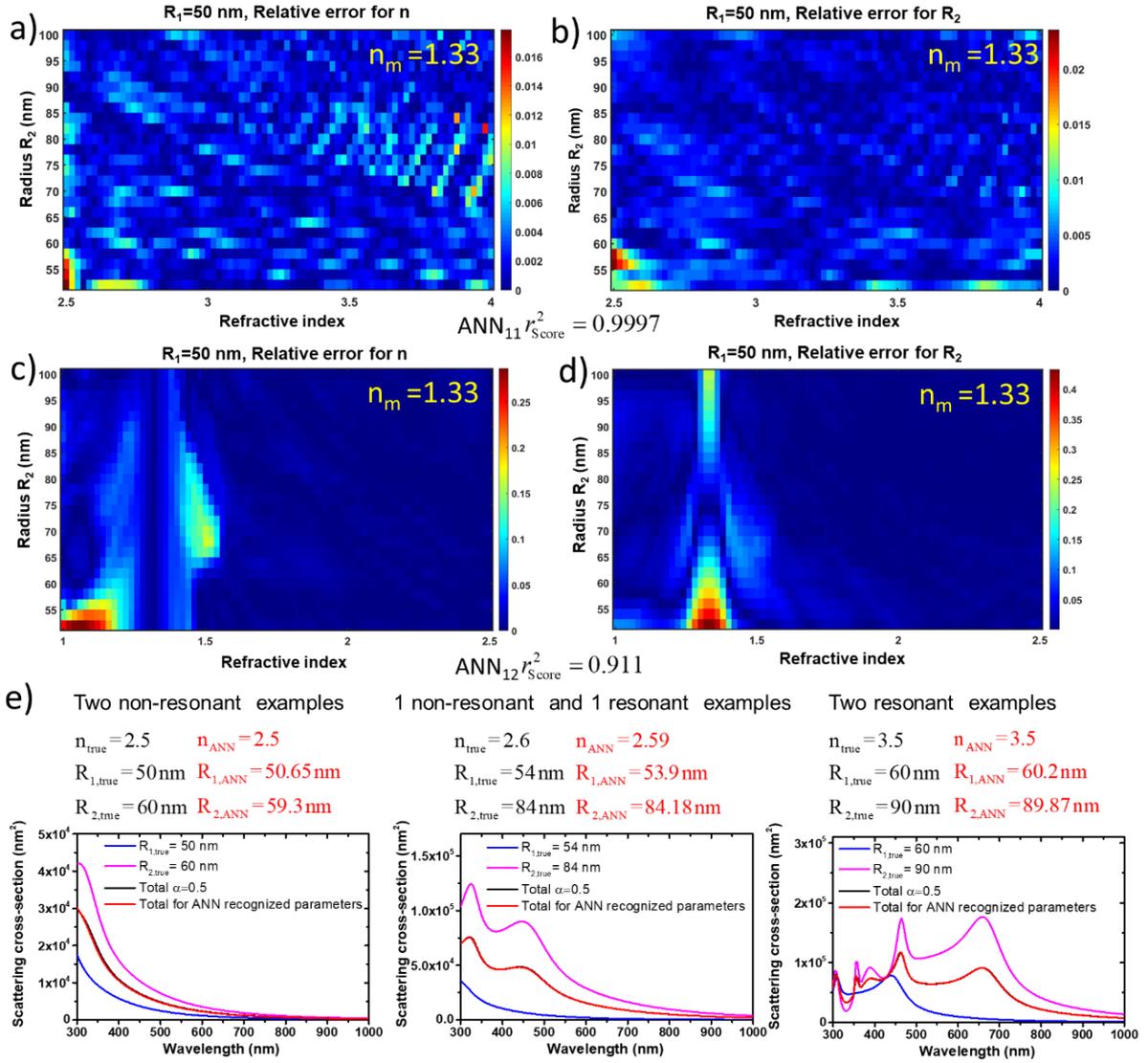

**Figure S5.** Three-parameter recognition scheme. The figure shows the relative-error maps for the refractive-index ($n$) and radius prediction ($R_2$) obtained from two trained ANNs for two intervals of refractive indexes: (a,b) $2.5 < n < 4$ and (c,d) $1 < n < 2.5$; the radii intervals are $50\,\text{nm} < R_1, R_2 < 100\,\text{nm}$. For the matrix RI, $n_m = 1.33$. To make these 2D maps, we fixed one of the parameters, e.g., the radius ($R_1$). (e) The examples of recognized and ground truth parameters of two different sizes of NSs ($R_1$ and $R_2$) of the same "$n$" material, $\alpha$ shows the relative concentration of NSs for three cases: (i) two non-resonant, (ii) one-resonant and one non-resonant, (iii) and two resonant NSs. Four curves [of scattering spectra] show two spectra of NSs of similar $\alpha$ concentration, the sum of these two spectra and the spectrum for recognized parameters (red curves). The resulting optical spectra based on the recognized parameters (red curves) show remarkable similarity with the ground truth spectra (black curves).

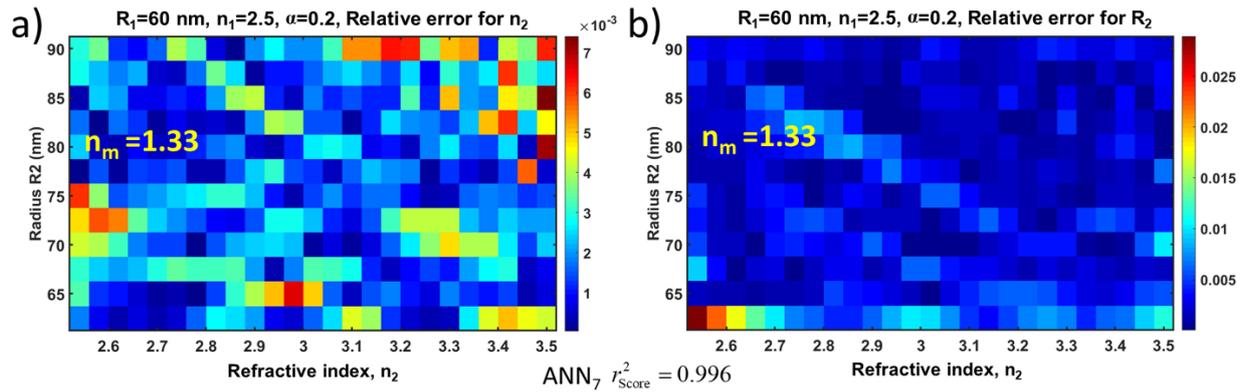

**Figure S6.** Five-parameter recognition scheme. The figure depicts the relative-error maps for the refractive index $(n_2)$ and the radius $(R_2)$ obtained from trained ANN7. Relative-error maps for (a) the refractive-index $(n_2)$ and (b) radius prediction $(R_2)$ obtained from recognized parameters by ANN7 for the intervals of refractive indexes $2.5 < n_1, n_2 < 3.5$ and for radii $60\,\text{nm} < R_1, R_2 < 90\,\text{nm}$. The interval of concentration coefficient is $0.2 < \alpha < 0.8$. The matrix RI is $n_m = 1.33$.

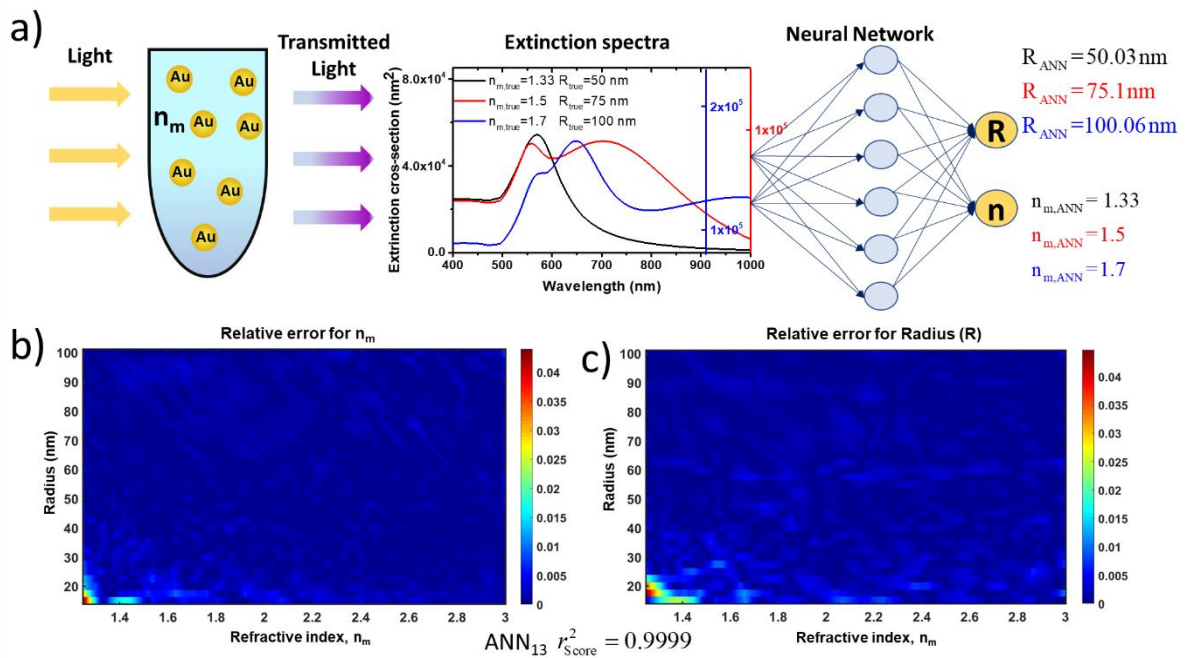

**Figure S7.** Two-parameter recognition scheme for the plasmonic sensing system. (a) Recognized sizes $(R)$ of the gold NS and the refractive indexes $(n_m)$ of the matrix by the neural network from the extinction spectra. (b,c) Relative-error maps of the RI of the matrix and the map of the radius errors. In these graphs, we used the ANN9 trained for the intervals $15\,\text{nm} < R < 100\,\text{nm}$ and $1.25 < n_m < 3$.

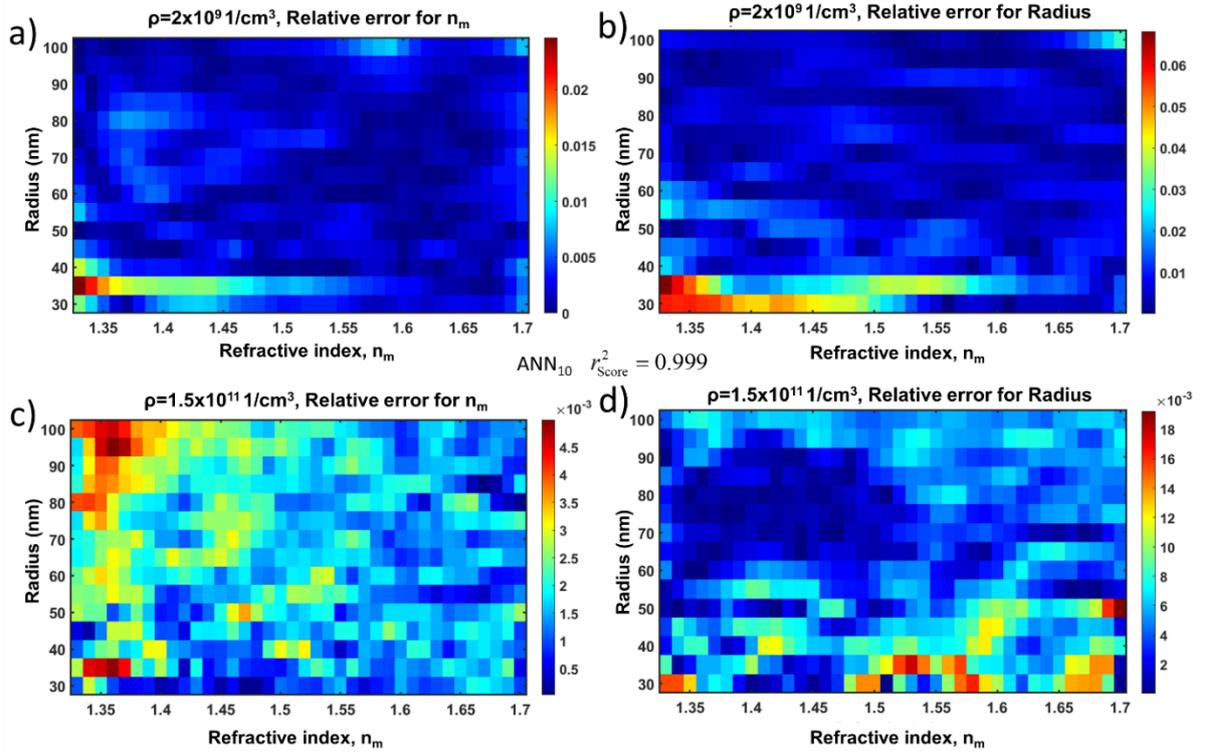

**Figure S8.** Three-parameter recognition model for the plasmonic sensing system, $(n_m, \rho, R)$. Relative-error maps for (a,c) the refractive-index of matrix $(n_m)$ and (b,d) density of gold NSs $(\rho)$ prediction obtained from the trained $ANN_{10}$ for the intervals of refractive indexes of matrix $1.33 < n_m < 1.7$, of radii $30\,\text{nm} < R < 100\,\text{nm}$, and of the concentration coefficient $3*10^8 < \rho < 3*10^{11}$. To make these 2D maps, we fixed one of the parameters, e.g., the density $\rho = 2*10^9\,1/\text{cm}^3$ for panel (a,b) and $\rho = 1.5*10^{11}\,1/\text{cm}^3$ for (c,d).

Figure S9 demonstrate the results for the case of plasmonic sensing considering a solution containing Au NSs of two different sizes $(30\,\text{nm} < R_1, R_2 < 100\,\text{nm})$ and diluted with different concentrations $(0.1 < \alpha < 0.9)$. The refractive-index interval is $1.33 < n_m < 1.7$. 2D maps are retrieved by fixing two parameters; e.g., we take $R_1 = 30\,\text{nm}$, $\alpha = 0.3$ and $R_1 = 30\,\text{nm}, \alpha = 0.7$ respectively for Figure S9a,b and S9c,d from the four-dimensional data $(R_1, R_2, \alpha, n_m)$ and obtain the relative errors for the remaining two parameters ($R_2$ and $n_m$).

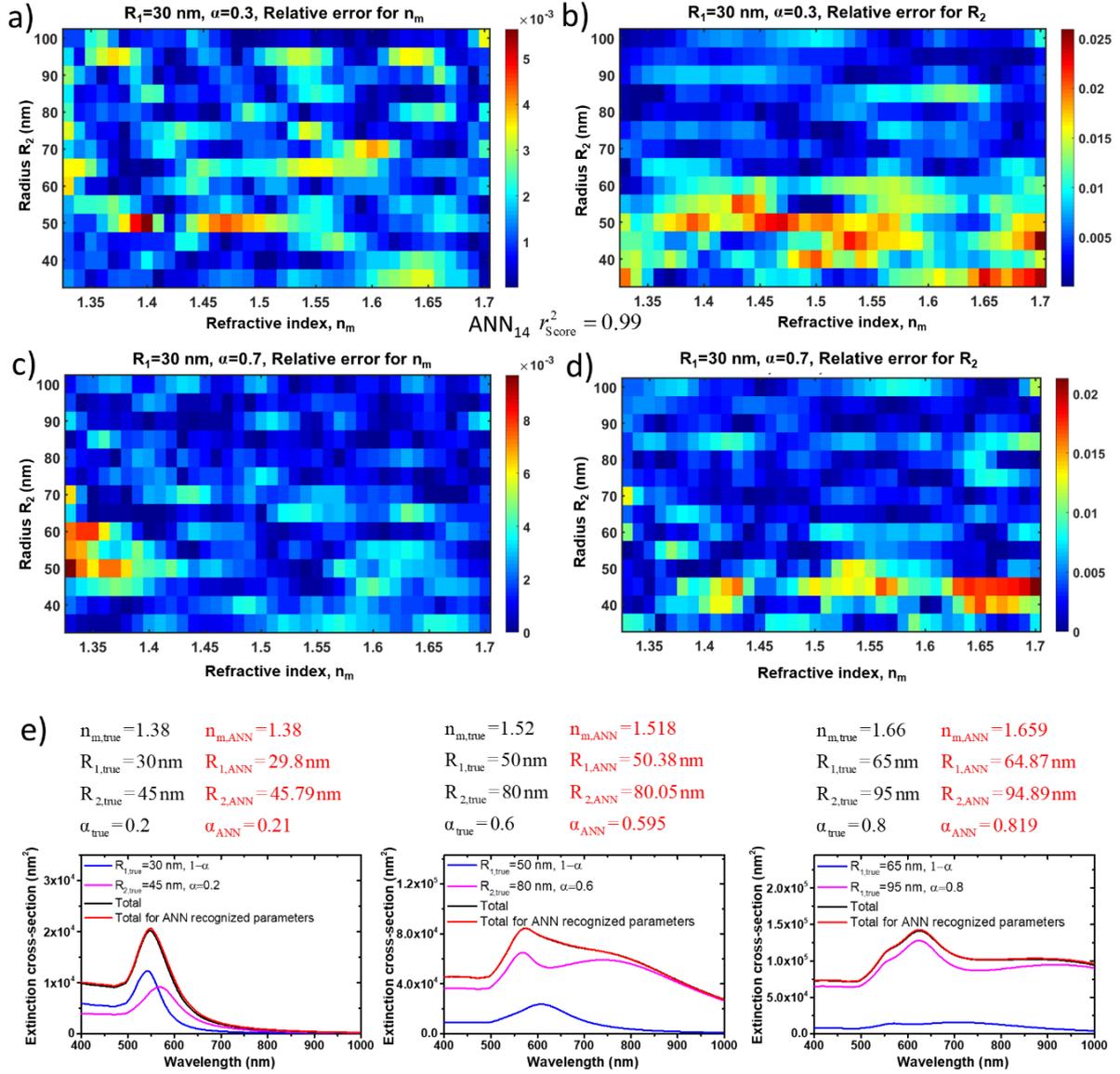

**Figure S9.** Four-parameter recognition model for the plasmonic sensing system. Relative-error maps for (a,c) the refractive-index of matrix ($n_m$) and (b,c) radius of gold NSs ($R_2$) prediction obtained from the trained ANN$_{13}$ for the intervals of refractive indexes of matrix $1.33 < n_m < 1.7$, of radii $30\,\text{nm} < R_1, R_2 < 100\,\text{nm}$, and of the concentration coefficient $0.1 < \alpha < 0.9$. (e) The examples of recognized and ground truth parameters of gold NSs of two different sizes ($R_1$ and $R_2$) of different $\alpha$ concentrations. Four curves of extinction spectra show for two NSs of $\alpha$ and $1-\alpha$ concentration, the sum of these two spectra and the spectrum for the recognized parameters by ANN$_{14}$. The refractive of the NSs is $n = n_{Au}$.

42